\documentclass[8pt,a4paper,twocolumn]{article}
\usepackage{graphicx,amsmath,color}
\usepackage{txfonts}
\usepackage{multirow}
\usepackage{subfigure}
\usepackage{natbib}
\usepackage{url}
\usepackage{authblk}

\newcommand{\C}[1] {\ensuremath{\times 10^{- #1}}}
\newcommand{\ld} {\ensuremath{\lambda/D \, }}

\bibpunct{(}{)}{;}{a}{}{,} 
   \title{Apodized phase mask coronagraphs for arbitrary apertures. II. Comprehensive review of solutions for the vortex coronagraph}

   \author[1,2]{A. Carlotti}
   \author[3]{L. Pueyo}
   \author[4]{D. Mawet}

   \affil[1]{Univ. Grenoble Alpes, IPAG, F-38000 Grenoble, France}
   \affil[2]{CNRS, IPAG, F-38000 Grenoble, France}
   \affil[3]{Space Telescope Science Institute, 3700 San Martin Drive, Baltimore, MD 21218, USA}
   \affil[4]{European Southern Observatory, Alonso de Cordova 3107, Vitacura, Santiago, Chile}

\begin{document}

\onecolumn
   \maketitle


   e-mail: \texttt{alexis.carlotti@obs.ujf-grenoble.fr}

 
  \abstract
   With a clear circular aperture, the vortex coronagraph perfectly cancels an on-axis point source and offers a 0.9 or 1.75 \ld inner working angle for topological charge 2 or 4, respectively. Current and near-future large telescopes are on-axis, however, and the diffraction effects of the central obscuration, and the secondary supports are strong enough to prevent the detection of companions $10^{-3}$-$10^{-5}$ as bright as, or fainter than, their host star. \\
   Recent advances show that a ring apodizer can restore the performance of this coronagraph by compensating for the diffraction effects of a circular central obscuration in a 1D modeling of the pupil. Our aim is to extend this work and design optimal apodizers for arbitrary apertures in 2D in order to tackle the diffraction effects of the spiders and other noncircular artefacts in the pupil. \\
   We fold this analytical result into a numerical optimization scheme that yields hybrid coronagraph designs that combine the advantages of the vortex coronagraph (small in IWA) and of shaped pupils coronagraphs (robustness to central obscuration and pupil asymmetric structures). The transmission of the apodizer is maximized, while constraints are set on the extremum values of the electric field that is computed in chosen regions of the Lyot plane through closed form expressions derived for even topological charges. Optimal apodizers are computed for topological charges 2 and 4 vortex coronagraphs and for telescope apertures with 10-30\% central obscurations and 0\%, 0.5\%, and 1\% thick spiders. 
   \\
   We put the results of our numerical optimizations in perspective with the analytical solutions and show that our apodizations converge to the ring apodizations. We then characterize the impacts of the obscuration ratio and the thickness of the spiders on the throughput and the IWA. For the apodized charge-2 vortex coronagraph the throughputs are slightly below those of the ring apodized vortex coronagraph, and the inner working angle is mostly unaffected by the apodization. The throughputs of the apodizers for the charge-4 vortex coronagraph are higher than those of the ring apodized vortex coronagraph. This effect increases with the obscuration ratio, though the inner working angle does, too, and it ranges between 2 and 3\ld. \\
   The results presented in this paper show that high contrast at small inner working angles can be obtained with a vortex coronagraph for on-axis telescopes, in spite of the presence of a secondary mirror and its secondary support structures.


%

\section{Introduction}


%
%


The spectral characterization of Earth-like planets around M, F, G, K stars at a few tens of parsecs from our Sun requires a $10^{-7}$ to $10^{-10}$ contrast at a few tens of milliarc-seconds (mas) from the host star in 20\% bandwidths. Self-luminous planets, which are only tens to hundred million years young, require a less demanding contrast to be imaged than older planets. Since the observation of the former must be preferably done in the infrared part of the spectrum, the chromatic scaling of the point-spread function (PSF) of the instrument partially compensates for this advantage, however.

On-axis 30-40m extremely large telescopes equipped with next-generation coronagraphs may provide the contrast, the resolution, and the large number of photons that are mandatory for ground-based observations. Few coronagraphs, however, can efficiently cope with the diffraction effects of the central obscuration and the secondary supports.


Adapting a coronagraph to an arbitrary aperture has been an intense research topic in the past few years. \cite{Soummer2009Arbitrary,Soummer2011} explain how the spheroidal prolate apodization of an apodized pupil Lyot coronagraph (APLC) can be adapted iteratively to a given aperture. \cite{Pueyo2013} study a similar problem, but explicitly constrains the contrast in chosen regions of the image plane, while in the previous case the high contrast was adjusted by varying the radius of the Lyot mask. Nonetheless, APLCs suffer from a rather large IWA.

Shaped pupils, which were initially optimized for high-contrast imaging in one dimension \citep{Spergel2001,Vanderbei2003,Kasdin2007}, can be numerically optimized in two dimensions (2D) for any telescope aperture \citep{Carlotti2011, Vanderbei2012}. Their versatility, robustness, and achromaticity make them good candidates for compact coronagraphic instruments: unlike APLCs, shaped pupils do not rely on a Lyot mask or a Lyot stop to create a high contrast, although a field stop is probably mandatory, given the limited dynamic range of detectors. Like APLCs, 2D-shaped pupil coronagraph have relatively large IWA, usually 3-5 \ld for apertures with obscurations of 10-30\% and for a $10^{-7}-10^{-10}$ contrast. The smallest IWA are usually obtained at the expense of the size of the discovery space.

All these coronagraphs require an extreme adaptive optics system to correct for the phase and the amplitude aberrations of the wavefront. This system can be composed of a single deformable mirror (DM), but corrections occur on only one side of the image plane \citep{Malbet1995}. As proposed in \cite{Pueyo2007} and demonstrated in \cite{Pueyo2009}, a system of two DMs in nonconjugate planes makes it possible to create symmetric dark holes in broadband. Such a system was recently test in JPL's high-contrast imaging testbed where it was used to obtain a 3$\times 10^{-10}$ in 10\% broadband \citep{Riggs2013}.

As suggested by \cite{Kasdin2011}, it appears that the wavefront control system can be actively used to create high contrast with arbitrary apertures, thus relaxing the specifications of the coronagraph. \cite{Pueyo2013} show how to use two DMs to spatially redistribute the energy density in the pupil plane so as to artificially decrease the size of the spiders and, for segmented telescopes, the gaps between the segments. \cite{Carlotti2013SPIEc} detail methods to optimize stroke commands to be sent to a DM to create a $10^{-6}-10^{-7}$ contrast with a centrally obscured aperture.

Pupil mapping is the core principle of the phase induced amplitude apodization (PIAA, \cite{Guyon2003}) technique. Combined with a complex amplitude focal plane mask (PIAACMC, \cite{Guyon2010}), it offers a very promising solution to the problem of small inner working angles as it results in much smaller IWA (down to 0.64\ld), while offering about 50\% throughput. The currently investigated manufacturing process of the focal plane mask involved in this instrumental concept remains challenging, however \citep{Newman2013}.


With a clear circular aperture of diameter $D$, looking at a wavelength $\lambda$, the four-quadrant phase mask coronagraph (4QPM, \cite{Rouan2000}) and the vortex coronagraph (VC, \cite{Mawet2005}) perfectly cancel an on-axis, unresolved point source, and detect companions as close as 0.9-1.75 \ld (0.9 for the 4QPM and a vortex with a topological charge 2, and 1.75 for a VC with a topological charge 4). 

The 4QPM coronagraph is a second order coronagraph: its off-axis transmission goes as $\theta^2$, where $\theta$ is the angular distance to the star. This makes the 4QPM coronagraph quite sensitive to jitter and to the finite size of the star. Like the 4QPM coronagraph, a VC with a topological charge 2 is a second order coronagraph. A charge 4 VC is a fourth order coronagraph, however: its off-axis transmission goes as $\theta^4$ instead of $\theta^2$. It is thus much less sensitive to the finite stellar size and to low order aberrations, which would otherwise limit the performance of the VC as they do with the 4QPM.

Another fourth order phase mask coronagraph is the eight-octant phase mask (8OPM, \cite{Murakami2008, Carlotti2009}), which is to the 4QPM, as the charge 4 VC is to the charge 2 VC. The 4QPM and the 8OPM attenuate off-axis sources along their phase edges, however, and this limits the extent of the discovery space space around the star, especially in the case of the 8OPM.

Nonetheless, the performance of the 4QPM coronagraph and the VC is greatly reduced when the telescope is on-axis. For instance numerical simulations predict a $10^{-4}-10^{-5}$ contrast between 1 and 5 \ld from the star, for a VC used with a 14\% centrally obscured aperture such as one of the very large telescope (VLT) 8m-class unit telescopes (UT). On-sky results have been obtained with a VC installed on VLT/NACO \citep{Mawet2013VLT}. Recently the VC has also been installed at two other telescopes: the Subaru telescope, and the Large Binocular Telescope.

A dual-stage coronagraph can be used to cancel the diffraction effects of a circular central obscuration with a 4QPM \citep{Galicher2011} or a VC \citep{Mawet2011}, but it can only partially attenuate those of the secondary supports (or any other artifacts in the pupil plane). It also requires twice as many components, making the alignment more difficult and increasing the size of the coronagraph.

It is possible to apodize the aperture to avoid the diffraction effects that prevent the 4QPM or the VC to create high-contrast at a small IWA with an on-axis telescope. 

As a matter of fact, a proof of concept for an apodized VC has already been successfully tested on the sky: a small, clear circular subaperture with a 1.5m diameter has been used at the Palomar telescope with a VC \citep{Serabyn2010Nat,Mawet2011APJL}. Unfortunately, this results in an effective resolution three times as small and a throughput eight times as small as what could have been obtained with the main telescope aperture.

It was shown in \cite{Carlotti2013} that shaped pupils can be optimally designed for a given combination of an arbitrarily-shaped aperture and a phase mask. Examples of such designs were numerically optimized for one of the 8m unit telescopes of the VLT, and for a 4QPM coronagraph, creating a few $10^{-10}$ contrast at 1 \ld (41mas in the H-band) from an unresolved star with a system throughput of about 64\%.

\cite{Mawet2013} shows that a ring-apodized vortex coronagraph (RAVC) can perfectly attenuate the on-axis light with a circular aperture with a circular central obscuration. Secondary supports are not taken into account, however. For a charge 2 VC this apodizer is composed of two rings: one is fully transmissive while the other is only partially transmissive. For a topological charge 4 vortex, a thin dark annulus separates the two previously described rings. The obscuration ratio of the aperture sets the radii of the rings and transmission of the outermost ring for which the throughput of the coronagraph is maximal.

The Lyot stop is directly constrained by the optimal rings radii. The same property characterizes the apodized 4QPM: the throughput of the coronagraph depends on the choice of the Lyot stop, namely the ratio of its central obscuration.

\begin{figure}[]
\centering
\includegraphics[width=0.9\textwidth]{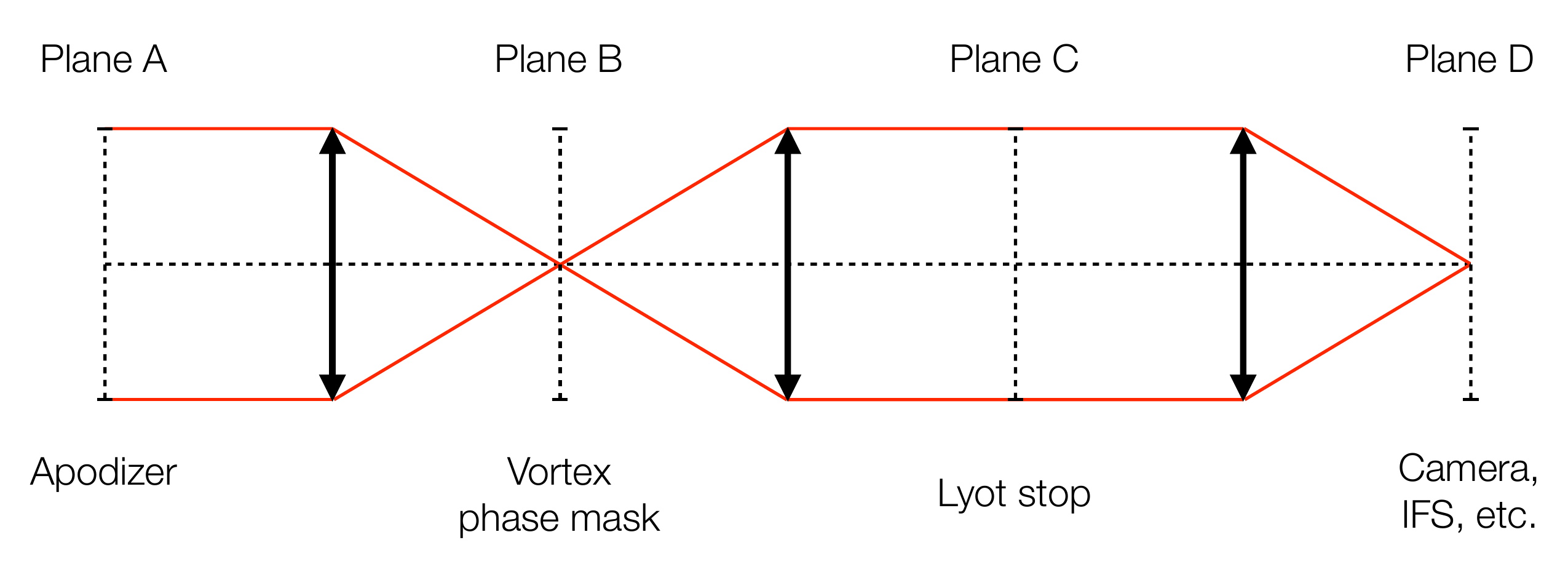}
\caption{Optical layout of an apodized vortex coronagraph. The apodizer, vortex phase mask, Lyot stop and camera are located in the successive pupil and image planes A, B, C, and D.}
\label{NewLayout}
\centering
\end{figure}

Fig. \ref{NewLayout} displays the optical layout of an apodized vortex coronagraph: an apodizer is located in a pupil plane A. In a subsequent image plane B, the Fourier transform of the electric field that goes through the apodizer is multiplied by the complex amplitude of the phase mask. In the reimaged pupil plane C, i.e., the Lyot plane, the Lyot stop blocks the diffracted on-axis light. Finally, the science camera could be located in a reimaged focal plane D.

Shaped pupils can be numerically optimized for the VC as they have already been for the 4QPM coronagraph, but the complexity of the computation process is much higher than in the case of the 4QPM. This is due to the fact that the pupil-to-pupil transform described in Eq.11 of \cite{Carlotti2013} cannot be written as nested sums. It must be directly evaluated as a 2D sum instead of 2 nested 1D sums.

As a consequence the memory requirements scale as the fourth power of the number of points chosen to discretize the pupil, while they only scale as the square for the case of the 4QPM. While this does not fundamentally preclude any computation, it makes the optimization of pupil apodizers much more demanding in terms of numerical resources, than their 4QPM counterparts  \citep{Carlotti2013}.


In this paper, we build upon and devise new methods to solve the limitations of the early papers \citep{Carlotti2013,Mawet2013} so as to design apodizers for 2D apertures and a VC.

A new computer dedicated to solving large scale optimization problems has made possible the computation of apodizers for the VC, with two-fold symmetry masks computed over $512^{2}$ points in a few hours. This computer has a four-core 3.6 Ghz processor and 64 GB of RAM. AMPL, a mathematical programming language is used to transcribe the optimization problem in a language that can be understood by one of the available solvers. In addition to the LOQO solver, the CPLEX and Gurobi solvers have been used. LOQO does not currently use more than one core of a processor, contrary to CPLEX and Gurobi, which can both use multiple cores. It should be noted that AMPL itself does not support parallel computing either.

To compute apodizers for the VC, we also had to modify the nature of the optimization problem. Indeed, there is a difference in the constraints set in the numerical optimization of apodizers for the VC and for the 4QPM coronagraph. Linear constraints were used for the 4QPM in \cite{Carlotti2013}. This was possible because, with an aperture with two axes of symmetry such as the VLT's, the 4QPM creates a purely real amplitude in the Lyot plane - and applying the constraints on the amplitude or the intensity does not make a difference. This is not the case with the VC: constraints must be set on the intensity of the electric field, which makes them quadratic.

We present in this paper the properties of apodizers optimized for a charge 2 VC and charge 4 VC with a 32\ld outer working angle, and designed to attenuate the on-axis light that goes through the Lyot stop by a $10^{6}$ factor, which result in a $10^{-8}-10^{-10}$ contrast at 1-3\ld from the star depending on the topological charge and the telescope aperture. An exhaustive number of apertures have been considered, with 10-30\% central obscurations, and for each of them 0, 0.5 and 1\% thick spiders. We compare the apodizers computed for spider-free apertures to the ring-apodizers, and we study the impact that an increasing spider thickness has on the performance metrics of the coronagraph.

Section \ref{Maths} details the mathematical formalism used to derive the closed form expressions of the pupil-to-pupil transforms used to compute the electric field in the Lyot plane. Section \ref{NOA} presents the optimization problem, and explains the methodology that is followed in the rest of the paper. Section \ref{Performance} details the throughput and the inner working angle of the numerical optimizations. For spiderless apertures it compares them with the results presented in \cite{Mawet2013}. It also addresses the limitations of broadband observations. Section \ref{Manufacturing} tackles the manufacturing aspects of this coronagraph. Section \ref{Conclusion} draws a conclusion to this paper.


\section{Analytical expressions for the apodized VC}\label{Maths}
\subsection{Formalism of apodized phase mask coronagraphs}
Here we briefly recall the formalism introduced in \cite{Carlotti2013} and extend it to the case of arbitrary charge vortex
The electric field in the image plane before the phase mask is a function of the transmission of the apodized aperture $A(x,y)$:
\begin{equation}\label{Eq1}
E(u,v)=\frac{e^{2i\pi F/\lambda}}{i \lambda F} \iint_{-D/2}^{D/2} A(x,y) \, e^{-2 i \pi (u x + v y)} \, dx \, dy.
\end{equation}
If $M(u,v)$ describes the phase mask, then the electric field right after the mask is written as the product of $E(u,v)$ and $M(u,v)$. The expression of the electric field in the subsequent pupil plane is the Fourier transform of that product:
\begin{equation}\label{Eq2}
P(\tilde{x},\tilde{y})= - i \lambda F e^{2i\pi F/\lambda} \iint_{-\infty}^{\infty} E(u,v) \, M(u,v) \, e^{2 i \pi (\tilde{x} u + \tilde{y} v)} \, du \, dv.
\end{equation}
Eq. \ref{Eq1} and \ref{Eq2} can be combined into a single expression:
\begin{equation}\label{Eq3}
P(\tilde{x},\tilde{y})= e^{4i\pi \frac{F}{\lambda}} \iint_{-\frac{D}{2}}^{\frac{D}{2}}\iint_{-\infty}^{\infty} A(x,y) M(u,v) e^{2 i \pi ((\tilde{x}-x) u + (\tilde{y}-y) v)} du dv dx dy.
\end{equation}
The Fourier transform $F(x,y)$ of the mask appears if Eq. \ref{Eq3} is written in a slightly different way:
\begin{equation}\label{Eq4}
\begin{split}
P(\tilde{x},\tilde{y}) &= e^{4i\pi F/\lambda} \iint_{-D/2}^{D/2} A(x,y) \, F(\tilde{x}-x,\tilde{y}-y) \, dx \, dy, \\
&\textrm{where } F(x,y) = \iint_{-\infty}^{\infty} M(u,v) \, e^{2 i \pi (x u + y v)} \, du \, dv.
\end{split}
\end{equation}
Eq. \ref{Eq4} can thus be seen as being proportional to the convolution product of the aperture $A(x,y)$ and the Fourier transform of the mask $F(x,y)$:
\begin{equation}\label{Eq5}
P(\tilde{x},\tilde{y})= e^{4i\pi F/\lambda} A(x,y) \ast F(x,y).
\end{equation}
If a closed form for the function $F(x,y)$ can be derived, then Eq. \ref{Eq5} can be used to compute the electric field in the Lyot plane without explicitly computing the electric field in the intermediate image plane.
The calculation of $F(x,y)$ depends on the chosen phase mask $M(u,v)$. It is more convenient to write the expression of $F(x,y)$ as a double integral with respect to $\rho$ and $\theta$, where $\rho$ and $\theta$ are the radial distance and the azymutal angle in the $(u,v)$ plane. For vortex phase masks, the function $M(\rho,\theta)$ does not depend on $\rho$, and equals $e^{i \theta l}$, where $l$ is the topological charge. This property makes it possible to proceed first with the integration with respect to $\rho$:
\begin{equation}\label{Eq6}
\begin{split}
F(x,y) &= \int_{0}^{2\pi} \int_{0}^{L} \rho \, M(\theta) \, e^{2 i \pi \rho (x \cos(\theta) + y \sin(\theta))} \, d\rho \, d\theta \\
         &= \int_{0}^{2\pi} M(\theta) \, \Psi(x \, \rho \cos(\theta) + y \, \rho \sin(\theta)) \, d\theta\\
\textrm{where } & \Psi(r)=\int_{0}^{L} \rho \, e^{2 i \pi \rho r} \, d\rho = \frac{(1-2 i \pi L r) \, e^{2 i \pi L r}-1}{4 \pi^2 r^2}.
\end{split}
\end{equation}
Note that $L$ represents the radius of the image plane mask (in units of \ld). While it is usually assumed that the mask is infinitely large, it is necessary to specify its size in order to derive the closed form expression for $\Psi(r)$ given in Eq.\ref{Eq6}.
\subsection{General expression for even topological charges}

In order to derive this result, we start by rewriting Eq.\ref{Eq6}, but instead of first integrating over $\rho$, we integrate first over $\theta$:

\begin{equation}\label{Eq6bis}
\begin{split}
F(x,y) &= \int_{0}^{2\pi} \int_{0}^{L} \rho \, e^{i l \theta} \, e^{2 i \pi (x \, \rho \cos(\theta) + y \, \rho \sin(\theta))} \, d\rho \, d\theta \\
&= \int_{0}^{2\pi} \int_{0}^{L} \rho \, e^{i l \theta} \, e^{2 i \pi r \rho \cos(\theta-\phi)} \, d\rho \, d\theta \\
&= \int_{0}^{L} \rho e^{i l \phi} \, \Big( \int_{0}^{2\pi} e^{i l \tilde{\theta}} \, e^{2 i \pi r \rho \cos{\tilde{\theta}}} \, d\tilde{\theta} \Big) \, d\rho,
\end{split}
\end{equation}

where we have first written $x=r \cos{\phi}$, $y=r\sin{\phi}$, and then proceeded with the change of variables $\tilde{\theta}=\theta-\phi$. We recognize here an expression which is closely related to the definition of the Bessel function of the $l$th order. The expression of $F(x,y)$ becomes:

\begin{equation}\label{Eq7bis}
\begin{split}
F(x,y) &= (-1)^{l} e^{i l \phi} \int_{0}^{L}  2 \pi J_{l}(2 \pi r \rho) \, \rho \, d\rho,
\end{split}
\end{equation}

and the final expression for $F(x,y)$ is obtained after integrating over $\rho$:

\begin{equation}\label{Eq8bis}
\begin{split}
F(x,y) &= e^{i l \phi} \frac{2 \pi L^2 (\pi L r)^{l}}{(2+l) \Gamma(l+1)} {}_{1}F_{2}(1+l/2 ; 2+l/2, 1+l ; - (\pi L r)^{2}),
\end{split}
\end{equation}

where $F$ is the generalized hypergeometric function.


\subsection{Expression for the first few even topological charges}

It is not difficult to derive for the first few even topological charge closed form expressions which only use Bessel functions.
The expression of $F(x,y)$ has previously been derived for the 4QPM and a VPM of topological charge $l=2$:
\begin{equation}\label{Eq7}
\begin{split}
F(x,y) = \frac{e^{2 i \phi(x,y)}}{\pi \, r(x,y)^2} \, \Big[ &-1 + J_{0}\big(2 \pi L \, r(x,y) \big) \\ 
&+\pi L \, r(x,y) \, J_{1}\big(2 \pi L \, r(x,y) \big) \Big]\\
\textrm{where } r(x,y) = \sqrt{x^2+y^2} & \textrm{ , and } \phi(x,y) = \tan^{-1}(y/x).
\end{split}
\end{equation}
In the case of a topological charge $l=4$, the integration over $\theta$ gives the following result:
\begin{equation}\label{Eq8}
\begin{split}
F(x,y) = \frac{e^{4 i \phi(x,y)}}{\pi \, r(x,y)^2} \times \, \Big[ 2 + 4 \, &J_{0}\big(2\pi L r(x,y)\big)\\ 
+ \Big(\pi \, L \, r(x,y)-\frac{6}{\pi \, L \, r(x,y)}\Big) \, &J_{1}\big(2 \pi L r(x,y)\big)\Big].
\end{split}
\end{equation}
And for a topological charge $l=6$, this expression becomes:
\begin{equation}\label{Eq9}
\begin{split}
F(x,y) = \frac{e^{6 i \phi(x,y)}}{\pi \, r(x,y)^2} \times \, \Big[ -3 + \Big( 9 - \frac{60}{\pi^2 \, L^2 \, r(x,y)^2} \Big) \, &J_{0}\big(2\pi L r(x,y)\big)\\ 
+ \Big(\pi \, L \, r(x,y)-\frac{36}{\pi \, L \, r(x,y)}+\frac{60}{\pi^3 \, L^3 \, r(x,y)^3}\Big) \, &J_{1}\big(2 \pi L r(x,y)\big)\Big].
\end{split}
\end{equation}
%
%

\section{Numerically optimized apodizers}\label{NOA}
Using Eq.\ref{Eq7} and Eq.\ref{Eq8} derived in the previous section, and proceeding with the same discretization process as described in \cite{Carlotti2013}, apodizers have been optimized for different apertures. The mask is discretized over N points along each axis of the pupil plane. We assume a unit aperture diameter, and the distance between nearest neighbors is $\Delta x = \Delta y = \frac{1}{N}$. 

\subsection{Set up of the optimization problem}\label{NOAZero}

The problem consists in maximizing the total transmission of the mask $\sum_{i=1}^{N} \sum_{j=1}^{N} A(x_{i},y_{j}) \Delta x \Delta y$ under the following constraints set on the extremum values of the transmission of the apodizer, and of the amplitude of the electric field $P(\tilde{x}_{k},\tilde{y}_{l})$ in the Lyot plane:
\begin{equation}\label{EqConstraint}
\begin{split}
 0 < A(x_{i},y_{j}) < 1 &  \textrm{   , with    } \{ x_{i}, y_{j} \} \in \Delta_{A} \\
|P(\tilde{x}_{k},\tilde{y}_{l})|^2 \le 10^{-c} & \textrm{   , with    } \{ \tilde{x}_{k}, \tilde{y}_{l} \} \in \Delta_{C},
\end{split}
\end{equation}
where $\Delta_{A}$ is the region defined by the telescope's aperture, $\Delta_{C}$ is the region defined by the Lyot stop, and $c$ measures the attenuation of the intensity in the reimaged pupil plane in a logarithmic scale. 


Since the vortex phase mask creates a complex amplitude in the Lyot plane, using linear constraints would force valid phasers in this plane to be located inside a square of size $\sqrt{2} \times 10^{-c/2}$. To explore all valid solutions, one must let the valid phasers live inside a larger area: a circle of radius $10^{-c/2}$ (which circumscribes the previous square). This  corresponds to the quadratic constraint that appears in Eq.\ref{EqConstraint}.

\subsection{Results in the case of noncircular apertures}\label{NonCircular}

Before studying exhaustively the properties of apodized vortex coronagraphs for various apertures, we first have to comment on the importance of the choice of the Lyot stop, and how the vortex phase mask constrains this choice. In particular, we want here to consider the case of noncircular apertures.

Many major ground-based and space-based telescopes have circular apertures (for instance the VLT's UTs, or the Hubble space telescope, among many others). The two Keck telescopes are segmented, however, and the shape of their outer edge is not circular (although their central obscuration is). The same can be said of the Gran Telescopio Canarias (GTC), which has a very similar pupil.

Other segmented telescopes are the future major gound-based telescopes such as the European extremely large telescope (E-ELT), the thirty-meter telescope (TMT), and a potentially large segmented space telescope. We should also mention the James Webb space telescope - whose primary mirror is also segmented - but  the design of its coronagraphic instruments has long been finalized, and no modification can be expected.

The formalism that we have introduced allows us to design apodizers for any type of aperture, and we have done so by considering the aperture of the Keck telescopes. Both primary mirrors are composed of 36 hexagonal segments of 1.8 meter (corner to corner).
The size of this aperture is close to 10.9 meters along its longest diameter, and 9m along its smallest diameter, and it is commongrayly accepted that the aperture mean diameter is 10m.

We have found that in the best case apodizers only have a 1-2\% transmission if the Lyot stop transmissive area encompasses the outermost segments (those that lie in the 9-11m diameter ring).

On the contrary, if the Lyot stop does not encompasses these outermost segments, i.e., if the Lyot stop outer edge is circular, then the transmission of the apodizers becomes significantly higher.

This is not too surprising: in its original design, the vortex coronagraph assumes a circular aperture. noncircular artifacts, such as the spiders or a noncircular outer edge, make it difficult to retrieve a high contrast. While an apodizer can be used to help restoring a high contrast, it cannot entirely compensate for the fact that the vortex phase mask works preferentially with circular clear aperture.

\begin{figure*}[]
\centering
\begin{tabular}{ccc}
\subfigure[]{\includegraphics[width=0.32\textwidth]{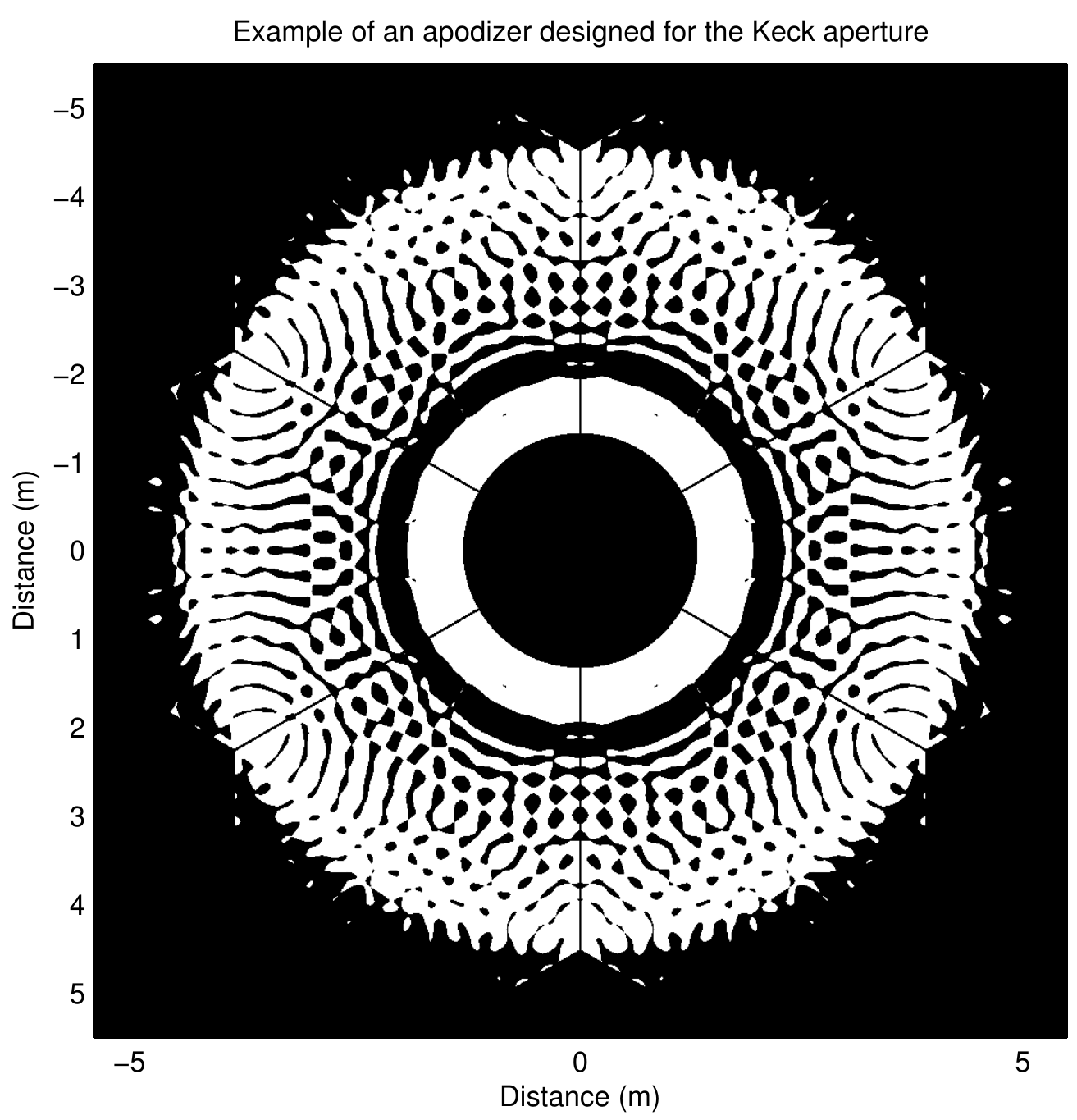}} \hspace{5mm}
\subfigure[]{\includegraphics[width=0.36\textwidth]{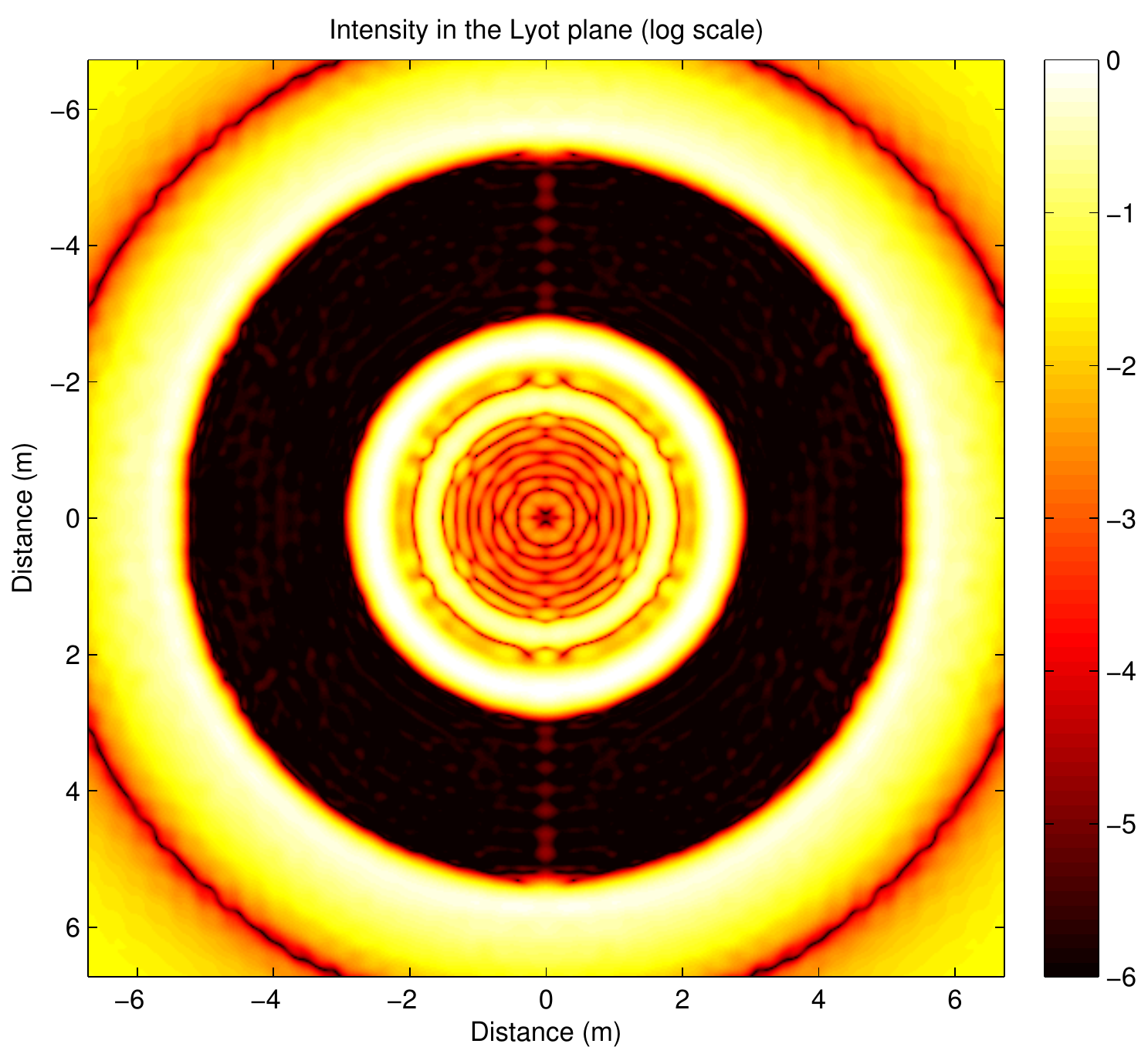}} \hspace{5mm}
\subfigure[]{\includegraphics[width=0.32\textwidth]{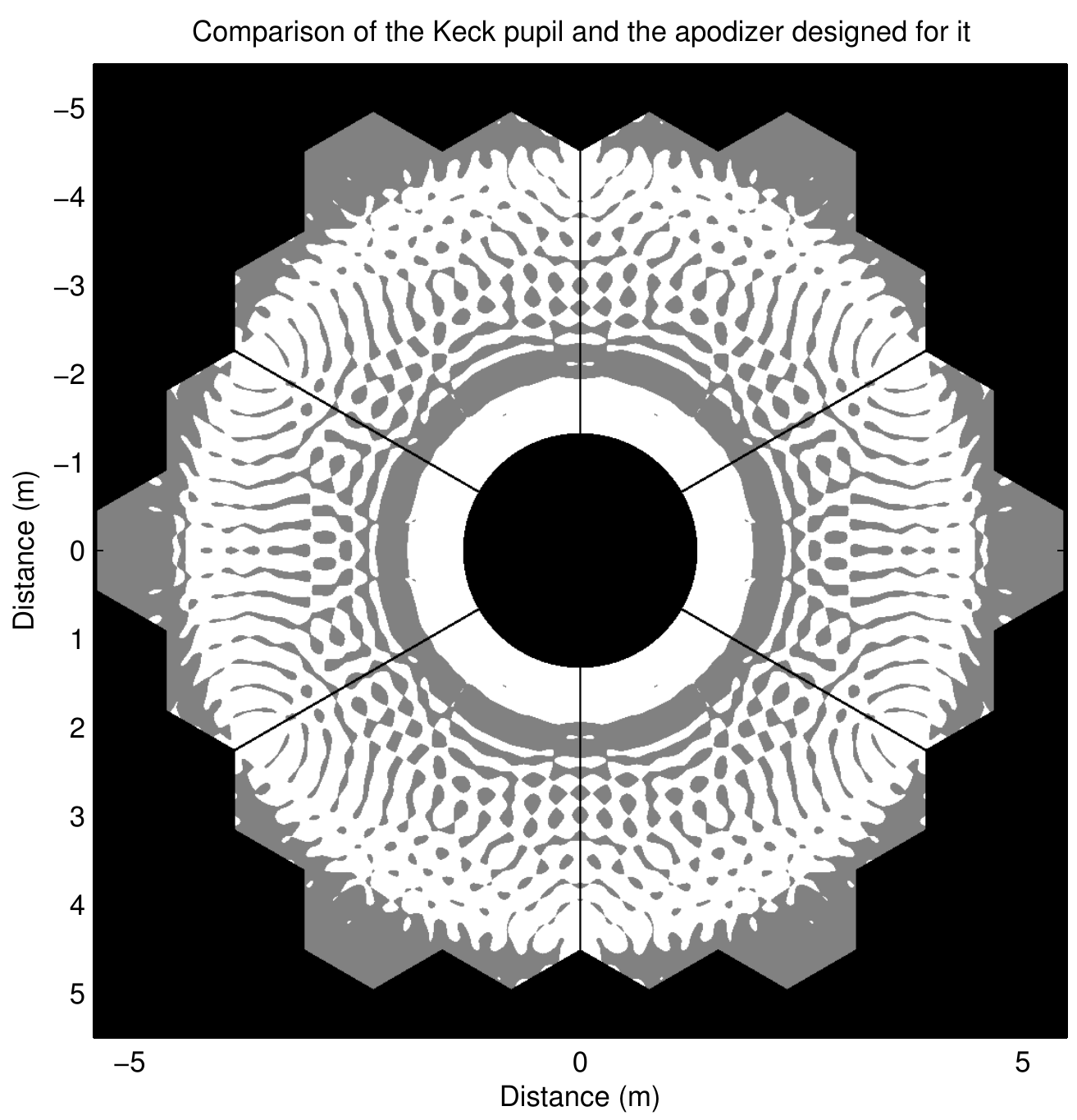}}
\end{tabular}
\caption{Example of a design for the Keck telescope, a charge 4 vortex coronagraph, and a circular Lyot stop: (a) apodizer transmission, (b) intensity in the Lyot plane (log scale), and (c) comparison of the Keck aperture and the apodizer. In the last figure the apodizer is represented in white, while the Keck aperture is represented in gray (wherever the apodizer transmission is zero).}
\label{Keck}
\centering
\end{figure*}

As it is shown in Fig.\ref{Keck}, the transmission of an apodizer computed for the Keck telescope and a circular Lyot stop is almost zero everywhere outside the circle sets by the Lyot stop's outer edge. No constrains was explicitly set on the apodizer, and the apodizer is only indirectly constrains by the Lyot stop properties.

With Keck, blocking light in the outermost segments means a maximum throughput 20\% smaller than the throughput associated with the entire pupil, and a mild resolution loss of about 8\% (assuming a 10m diameter for the aperture).

The apertures of the E-ELT and the TMT are composed of a much higher number of segments, however, and making the outer edge circular for them does not come with the same throughput or resolution limitation. Besides, in the case of E-ELT, the secondary mirror will be circular. Its diameter will be such that it will only capture the inner, largest circular area possible, making the effective diameter of the telescope 37m.

Since it appears necessary to restrict the Lyot stop's outer edge to a circle, and that it indirectly constrains the apodizer transmission to be zero outside the same circle, we have chosen to only consider circular apertures and circular Lyot stop in the rest of this study.

\subsection{Results in the case of circular apertures}\label{Circular}

Five linear obscuration ratios of 10, 15, 20, 25, and 30\% have been considered. The secondary support structure is orthogonal and their thickness is either 0, 0.5, or 1\%. The first case corresponds to a spider-free centrally obscured aperture, which is useful to consider when comparing the properties of the numerically optimized apodizers with those of the ring-apodizer which too are designed for an aperture with a central obscuration but no spiders. The 0.5 and 1\% thicknesses have been chosen to match the thickness of current telescopes' spiders. For instance the spiders of the Hale telescope at the Palomar observatory are 0.25\% thick, and those of the 8m unit telescopes at the VLT are 0.5\% thick.

Overall, the Lyot stops' geometry and the apertures' geometry are similar. The diameter of the Lyot stops is 96\% that of the aperture diameter. This is meant to prevent the diffraction effects of the finite size of the vortex phase mask. For the same reason, the secondary supports of the Lyot stops are twice as large as those of the apertures. The main difference between the Lyot stops and the apertures is the size of the central obscuration of the Lyot stops, which is larger for the Lyot stops than for the apertures, and is a key parameter that must be adjusted when looking for an optimal apodizer.



The two axes of symmetry displayed by the apertures and the Lyot stops are used to reduce the memory required during the optimizations. The transmission of the apodizers was optimized in one quadrant of the pupil plane over N=256 points (along each axis). The amplitude was computed in one quadrant of the Lyot plane over M=64 points. As the vortex mask was assumed to be 32\ld in diameter, this number of points is enough to satisfy the Nyquist-Shannon sampling theorem.

Note that for this number of points up to a third of the available memory was used. While further increasing it is possible, it was enough to assess the interest of the method by computing a large number of apodizers.


%

\section{Performance}\label{Performance}

\subsection{System throughput}

As it was observed with the apodized 4QPM coronagraph and with the ring-apodized VC, the throughput of the system depends on the size of the central obscuration of the Lyot stop: an optimal value exists for which the throughput is maximized.

\subsubsection{Comparison with the ring apodizers}

The ring-apodized vortex coronagraph proposed in \cite{Mawet2013} creates high-contrast in spite an aperture with a circular, central obscuration. The ring apodizers are formed of 2 or more rings with different transmissions. For instance, a charge 2 vortex requires a first ring with a $t=1$ transmission, surrounded by a $0<t<1$ second ring. In the case of a charge 4 vortex, a $t=1$ first ring is surrounded by a $t=0$ second ring, which itself is encircled by a $0<t<1$ third ring. The transmissions and radii of the rings depend on the obscuration ratio of the aperture.

Ring apodizers offer an elegant and relatively simple approach to a complicated problem, but they do not take the diffraction effects of the secondary supports into account. Moreover both the central obscuration and the shape of the pupil are assumed to be circular, which is not the case of all telescopes, although we showed that numerically optimized apodizers designed for noncircular apertures were forced to be quasi-circular by the VC.

The numerically optimized apodizers and the ring apodizers are designed to solve the same general problem, however, and it is interesting to try and compare how similar or not they can be. 

A first element of comparison comes from looking at the optimal apodizers that have been computed for the spider-free aperture. These apodizers look very much like the ring apodizers. The shaped pupils can also be divided into two or three areas, the only difference being that the transmission of the outer annulus is binary with the shaped pupils, whereas it has a gray transmission everywhere in the annulus of the ring-apodizers.

\begin{figure}[]
\centering
\begin{tabular}{c}
\subfigure[]{\includegraphics[width=0.45\textwidth]{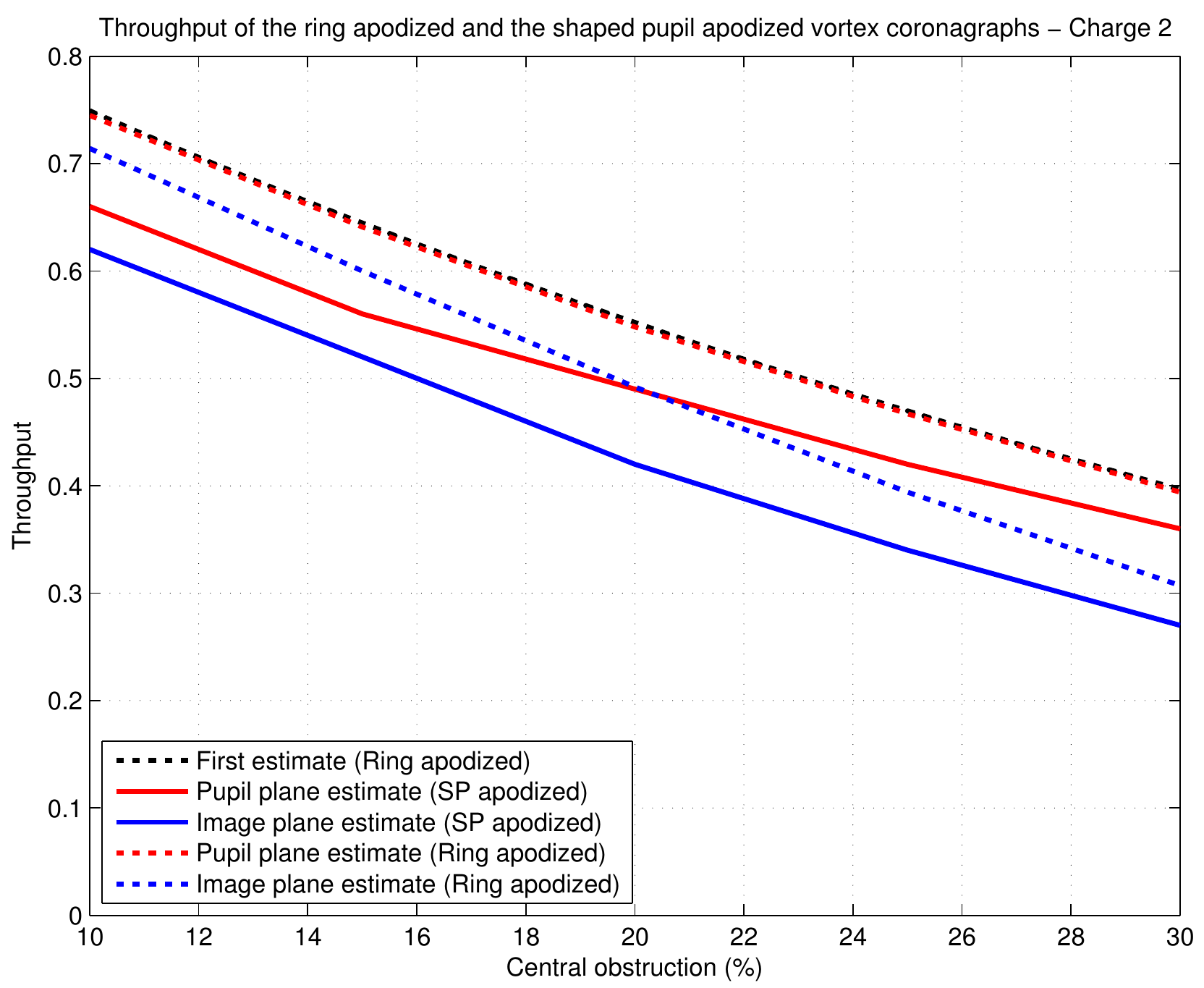}}
\subfigure[]{\includegraphics[width=0.45\textwidth]{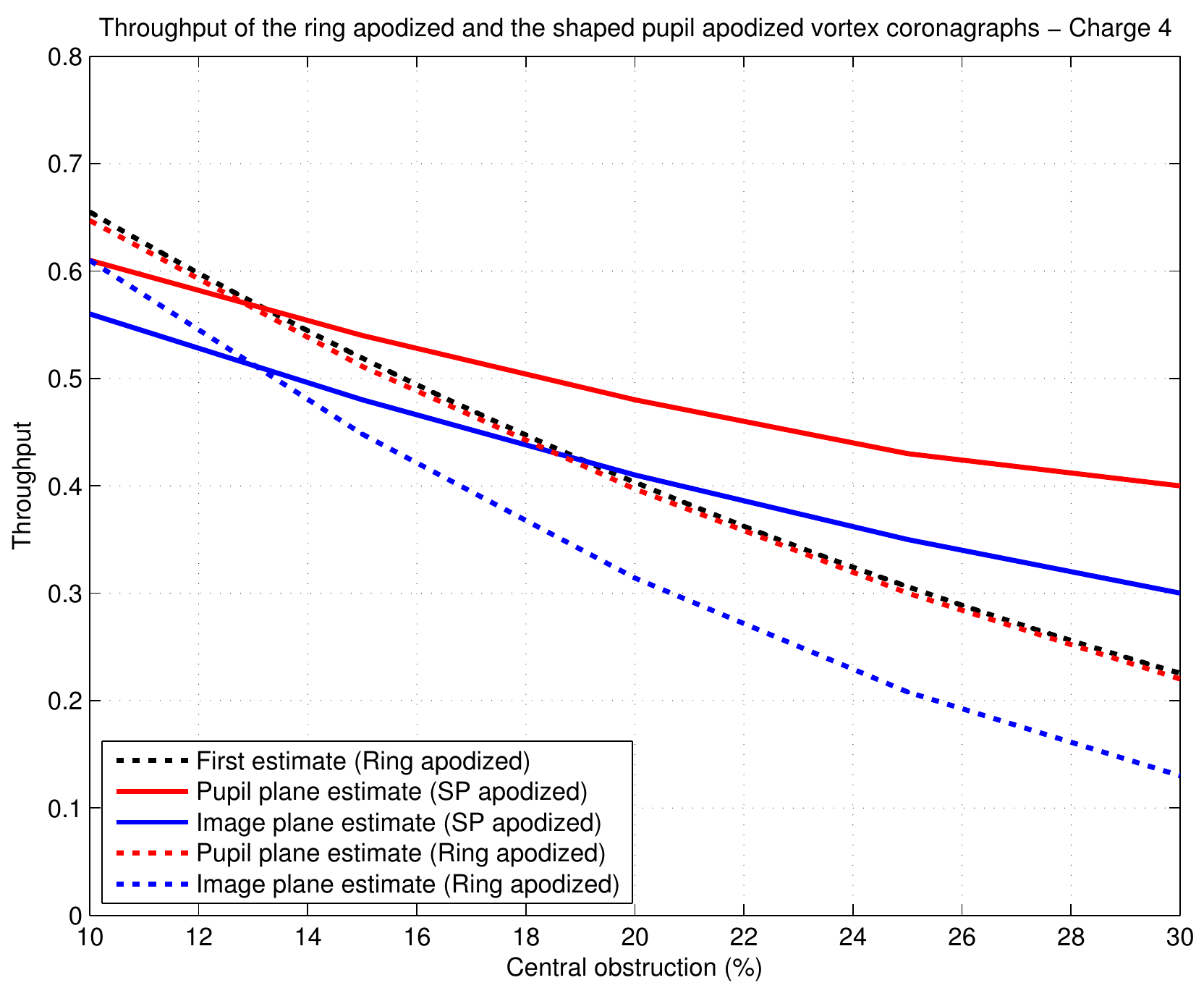}}
\end{tabular}
\caption{Comparison of the throughputs estimates of the ring apodized VC (dotted lines) and the shaped pupil apodized VC (solid lines) for topological charges 2 and 4. Different measures of the throughput are illustrated. Pupil plane estimates are showed in red, while image plane estimates are showed in blue. The dotted black line represents the throughputs given in \cite{Mawet2013}.}
\label{RingApodizerComparison}
\centering
\end{figure}

Fig.\ref{RingApodizerComparison} shows how the system's throughput of the ring apodizers change with the obscuration ratio of the aperture. The corresponding quantities for the optimal apodizers computed for the five 10-30\% central obscuration spider-free apertures are displayed for comparison.

Two types of throughput estimates are distinguished: pupil-plane estimates and image-plane estimates. Pupil-plane estimates are the ratio of the total planet intensity going through the Lyot stop, and the total planet intensity going through the plane located immediately before the apodizer. Image-plane estimates are the ratio of the total planet intensities measured in a small circular area centered on the planet with and without the apodized coronagraph. We chose this circular area to be 2 $\lambda/D$ in radius.

For comparison, the throughputs given in \cite{Mawet2013} are computed in the pupil-plane, although this computation was not performed by propagating light in a simulated system, but rather by relying on the analytic description of the apodized coronagraph.

A first observation is that the throughputs given in \cite{Mawet2013} for charges $l=2$ and $l=4$ are almost identical to the pupil-plane estimates we have computed by propagating a planetary companion in a simulated coronagraph. The very small differences that were observed are very likely due to the fact that the size of the VC had to be finite, and was set to 256 $\lambda/D$.

In the rest of this comparison analysis, we refer to the system throughput as the image-plane estimate of the throughput.

In the case of the charge $l=2$ VC, the system throughput associated with the numerically optimized apodizers is always lower than what it is with the ring apodizers. The difference gets smaller with the central obscuration, however. The system throughput for the ring-apodizers is 72\% for a 10\% central obscurations $CO_{A}$, and 30\% for $CO_{A}$=30\%, and this decrease is quite linear. As for the system throughput for the numerically optimized apodizers, it is 62\% for $CO_{A}$=10\%, and 27\% for $CO_{A}$=30\%. It is thus 10\% smaller for the smallest obscuration, and 3\% smaller for the largest obscuration.

On the other hand, in the case of the charge $l=4$ VC, the system throughput associated with the numerically optimized apodizers is smaller than that of the RAVC for very small obscurations, but it is significantly larger for large central obscurations. This difference goes from being -4\% for $CO_{A}$=10\%, to being 0 for $CO_{A}$=13\%, to being +10\% for $CO_{A}$=20\%, and +17\% for $CO_{A}$=30\%. For $CO_{A}$=10\% it is 56\%. It is 41\% for $CO_{A}$=20\%, and it is 30\% for $CO_{A}$=30\%.


The fact that (a) the focal plane masks have a finite size (32 \ld), and (b) the remaining on-axis energy inside the Lyot stop is small but not zero may be the main reasons that explain why the throughput of the apodized charge 4 VC is higher than it could have been expected. This is outside the scope of this paper, but we intend to address this point in future work. One of the difficulty preventing us to present results on this topic in this paper comes from the current impossibility to compute apodizers for larger OWA than 32\ld, as it requires a RAM quantity larger than what is available with the computers we have access to at the moment.


We already mentioned that, when looking for the values of $CO_{LS}$, we started with Lyot stops prescribed using the analytical expression of the ring apodizers and found a local optimum in that neighborhood. While not showed in Fig.\ref{RingApodizerComparison}, we have found that the $CO_{LS}$ values are very close to their counterpart for the ring apodizers for $l=2$, and close to but increasingly smaller for $l=4$.

This confirms that the formalism used to design ring apodizers can and should be used to infer a first estimate of the obscuration ratio of the Lyot stop used in the numerical optimizations.


\subsubsection{Additional impact of the spider thickness}

To find the optimal obscuration ratio, we have computed apodizers for several Lyot stops, starting with the obscuration ratio predicted by the closed form expressions derived in \cite{Mawet2013}, and iterating until a maximum was reached.


Tab. \ref{tabZERO} and tab. \ref{tabONE} list - for the topological charges $l=2$ and $l=4$ - the parameters of the optimal Lyot stops and of the apodizers that correspond to them, for the five apertures and the three spider thicknesses that were considered. To reach the maximum throughput, larger central obscurations $CO_{A}$ require larger Lyot stop obscurations $CO_{LS}$, and this decreases both the transmission of the apodizer $T_{A}$ and the maximum throughput of the coronagraph $T_{max}$.

The tables also display the transmission of the Lyot stop $T_{LS}$. It should be noted that the product of $T_{A}$ and $T_{LS}$ - which could be seen as the expected throughput of the coronagraph - is about 10\% smaller than $T_{max}$: some of the off-axis light gets diffracted outside the Lyot stop, and is thus blocked by it.

\begin{table}[]
\begin{center}
\begin{tabular}{|c|c|c|c|c|c||c|c|c|}
\hline $CO_{A}$ & $t_{S}$ & $CO_{LS}$ & $T_{A}$ & $T_{LS}$ & $T_{LP}$ & $T_{max}$ & IWA & Contrast \\ \hline \hline
10 & 0 & 32 & 90 & 83 & 66 & 62 & 0.9 & $5 \times 10^{-8}$ \\
15 & 0 & 38 & 85 & 80 & 56 & 52 & 0.9 & $4 \times 10^{-8}$   \\
20 & 0 & 50 & 87 & 70 & 49 & 42 & 0.8 & $5 \times 10^{-9}$   \\
25 & 0 & 52 & 80 & 69 & 42 & 34 & 0.9 & $6 \times 10^{-8}$   \\
30 & 0 & 58 & 79 & 64 & 36 & 27 & 0.9 & $8 \times 10^{-8}$   \\ \hline \hline
10 & 0.5 & 28 & 83 & 84 & 57 & 52 & 0.9 & $3 \times 10^{-8}$   \\
15 & 0.5 & 40 & 84 & 77 & 52 & 46 & 0.9 & $1 \times 10^{-8}$   \\
20 & 0.5 & 48 & 83 & 71 & 46 & 39 & 0.9 & $2 \times 10^{-8}$   \\
25 & 0.5 & 58 & 84 & 62 & 39 & 29 & 0.8 & $5 \times 10^{-9}$   \\
30 & 0.5 & 58 & 77 & 64 & 34 & 24 & 0.9 & $7 \times 10^{-9}$   \\ \hline \hline
10 & 1 & 28 & 63 & 84 & 38 & 29 & 1.1 & $3 \times 10^{-8}$   \\
15 & 1 & 40 & 64 & 77 & 35 & 26 & 1.1 & $5 \times 10^{-8}$  \\
20 & 1 & 48 & 65 & 71 & 32 & 23 & 1.2 & $6 \times 10^{-8}$   \\
25 & 1 & 58 & 67 & 62 & 28 & 18 & 1.5 & $7 \times 10^{-8}$   \\
30 & 1 & 58 & 67 & 64 & 29 & 18 & 1.4 & $2 \times 10^{-8}$   \\ \hline
\end{tabular}
\end{center}
\caption{Charge 2 VC: main parameters for the combination of apodizers and Lyot stops that give the highest system throughputs for the five aperture diameters $CO_{A}$ and the three spider thicknesses $t_{S}$. $CO_{LS}$ refers to the central obscuration of the Lyot stop. These three parameters are given in percent: of pupil diameter. The transmission of the apodizer and the Lyot stop are noted $T_{A}$ and $T_{LS}$ (in \% of incoming light). The maximum fraction of the incoming planet light transmitted by the Lyot plane is noted $T_{LP}$, and is given in \% of incoming light as well. The last three columns display the effective maximum throughput $T_{max}$ (measured in the image plane), the IWA (in units of \ld), and the mean contrast at one IWA from the star.}
\label{tabZERO}
\end{table}

Increasing $CO_{A}$ from 10 to 30\% decreases $T_{max}$ by about a third for the charge 4 AVC, no matter the spider thickness. For the charge 2 AVC, $T_{max}$ decreases by  50\% for the spider-less apertures, by 40\% for $t_{S}=0.5\%$, and by 23\% for $t_{S}=1\%$. 

Given the same central obscuration, thicker spiders, however, greatly decrease the coronagraphic throughput. The impact of the spider thickness $t_{S}$ on the throughput $T_{max}$ increases with $t_{S}$: while there is only a 5-8\% decrease of the throughput between coronagraphs designed for a spider-free aperture and designs with 0.5\% thick spiders, there is a 14-19\% decrease of the throughput between the latter and 1\% thick spiders. The smaller the central obscuration of the aperture, the larger the decrease.

\begin{table}[]
\begin{center}
\begin{tabular}{|c|c|c|c|c|c||c|c|c|}
\hline $CO_{A}$ & $t_{S}$ & $CO_{LS}$ & $T_{A}$ & $T_{LS}$ & $T_{LP}$ & $T_{max}$ & IWA & Contrast \\ \hline \hline
10 & 0 & 38 & 88 & 79& 61 & 56 & 1.8 & $2 \times 10^{-10}$ \\
15 & 0 & 44 & 85 & 74 & 54 & 48 & 2.3 & $2 \times 10^{-10}$   \\
20 & 0 & 50 & 83 & 70 & 48 & 41 & 2.6 & $3 \times 10^{-10}$   \\
25 & 0 & 50 & 76 & 72 & 43 & 35 & 2.8 & $3 \times 10^{-10}$   \\
30 & 0 & 56 & 75 & 67 & 40 & 30 & 3.2 & $6 \times 10^{-10}$   \\ \hline \hline
10 & 0.5 & 36 & 84 & 81 & 56 & 50 & 1.8 & $2 \times 10^{-9}$   \\
15 & 0.5 & 42 & 81 & 77 & 49 & 42 & 2.3 & $6 \times 10^{-10}$   \\
20 & 0.5 & 50 & 79 & 71 & 43 & 35 & 2.5 & $5 \times 10^{-9}$   \\
25 & 0.5 & 50 & 73 & 72 & 39 & 31 & 2.8 & $4 \times 10^{-9}$   \\
30 & 0.5 & 56 & 71 & 67 & 35 & 25 & 3.2 & $7 \times 10^{-9}$   \\ \hline \hline
10 & 1 & 32 & 63 & 85 & 37 & 28 & 2.2 & $2 \times 10^{-8}$   \\
15 & 1 & 42 & 64 & 78 & 34 & 25 & 2.3 & $2 \times 10^{-8}$  \\
20 & 1 & 50 & 63 & 72 & 29 & 21 & 2.5 & $1 \times 10^{-8}$   \\
25 & 1 & 50 & 57 & 73 & 25 & 18 & 2.9 & $2 \times 10^{-8}$   \\
30 & 1 & 56 & 55 & 68 & 22 & 14 & 3.2 & $3 \times 10^{-8}$   \\ \hline
\end{tabular}
\end{center}
\caption{Charge 4 VC: main parameters for the combination of apodizers and Lyot stops that give the highest system throughputs for the five aperture diameters $CO_{A}$ and the three spider thicknesses $t_{S}$. $CO_{LS}$ refers to the central obscuration of the Lyot stop. These three parameters are given in percent: of pupil diameter. The transmission of the apodizer and the Lyot stop are noted $T_{A}$ and $T_{LS}$ (in \% of incoming light). The maximum fraction of the incoming planet light transmitted by the Lyot plane is noted $T_{LP}$, and is given in \% of incoming light as well. The last three columns display the effective maximum throughput $T_{max}$ (measured in the image plane), the IWA (in units of \ld), and the mean contrast at one IWA from the star.}
\label{tabONE}
\end{table}

\begin{figure}[]
\centering
\begin{tabular}{ccc}
\subfigure[]{\includegraphics[width=0.3\textwidth]{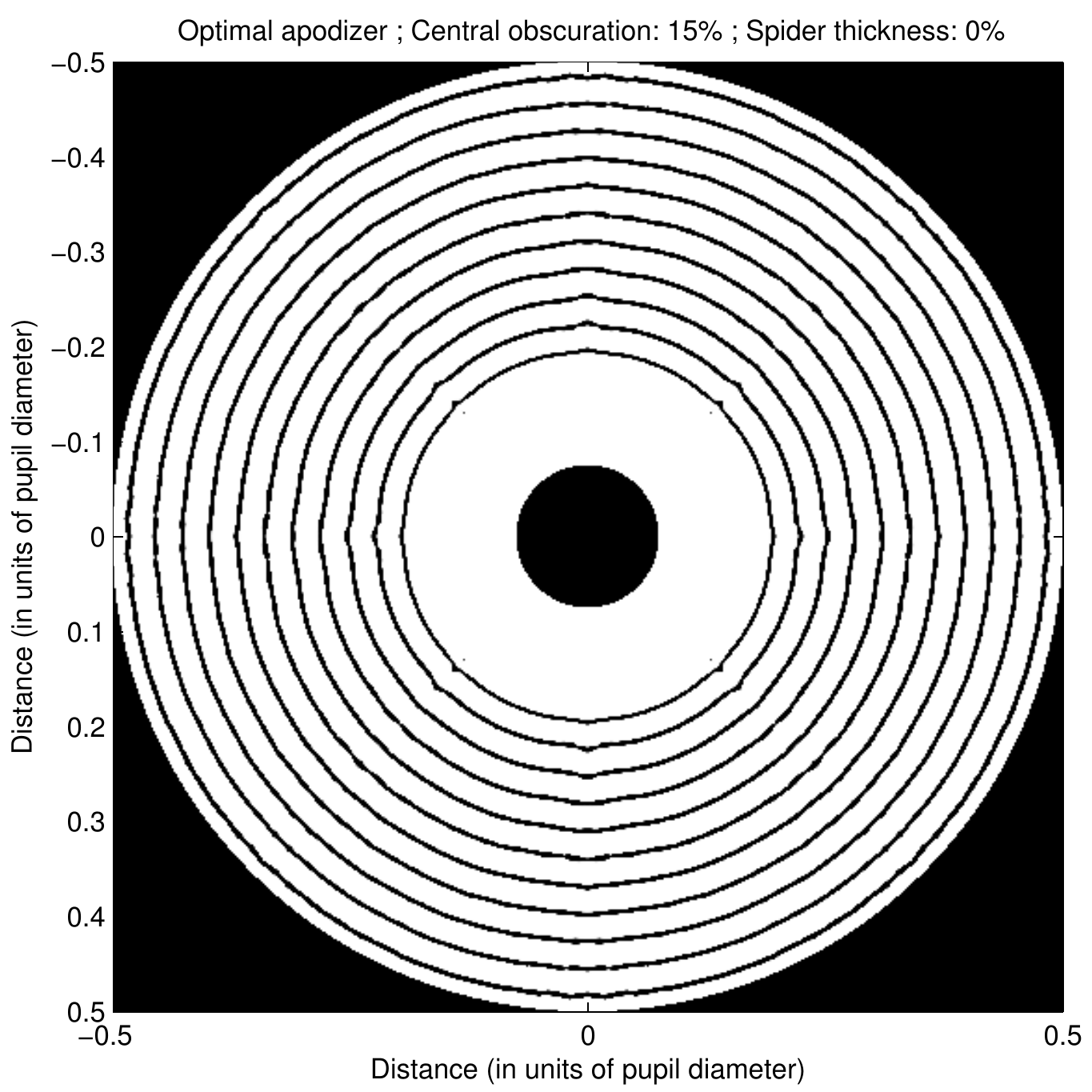}}
\subfigure[]{\includegraphics[width=0.3\textwidth]{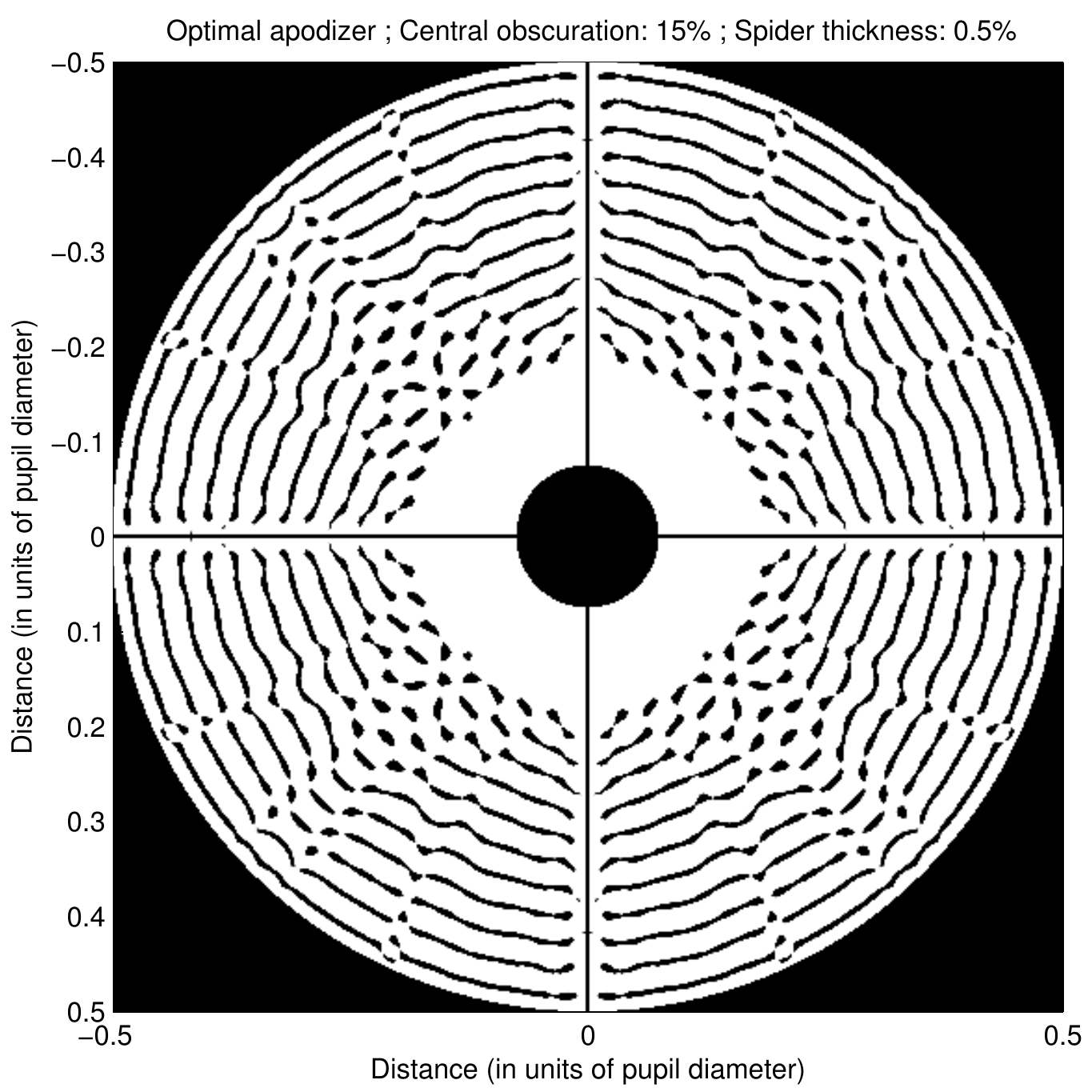}} 
\subfigure[]{\includegraphics[width=0.3\textwidth]{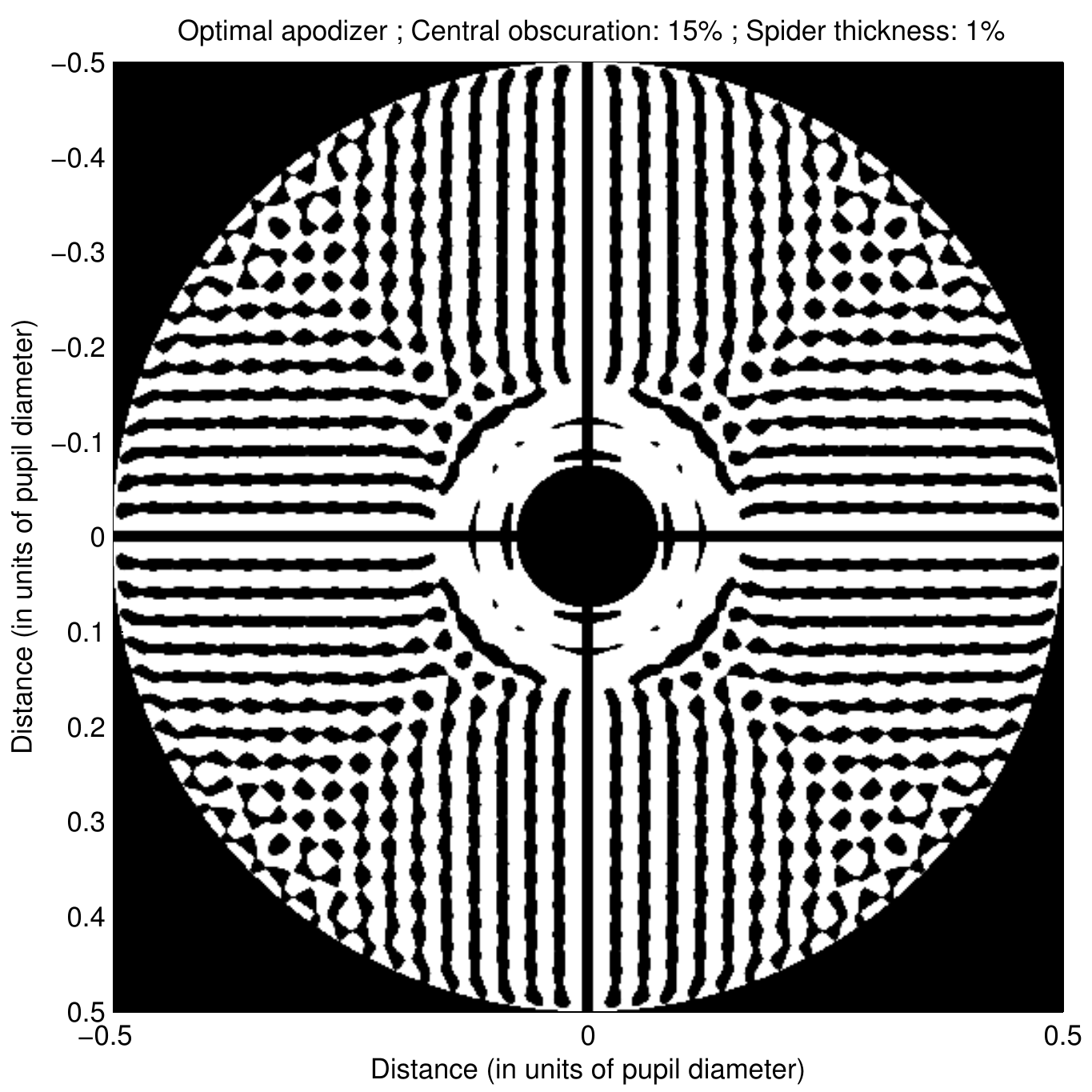}} \\
\subfigure[]{\includegraphics[width=0.3\textwidth]{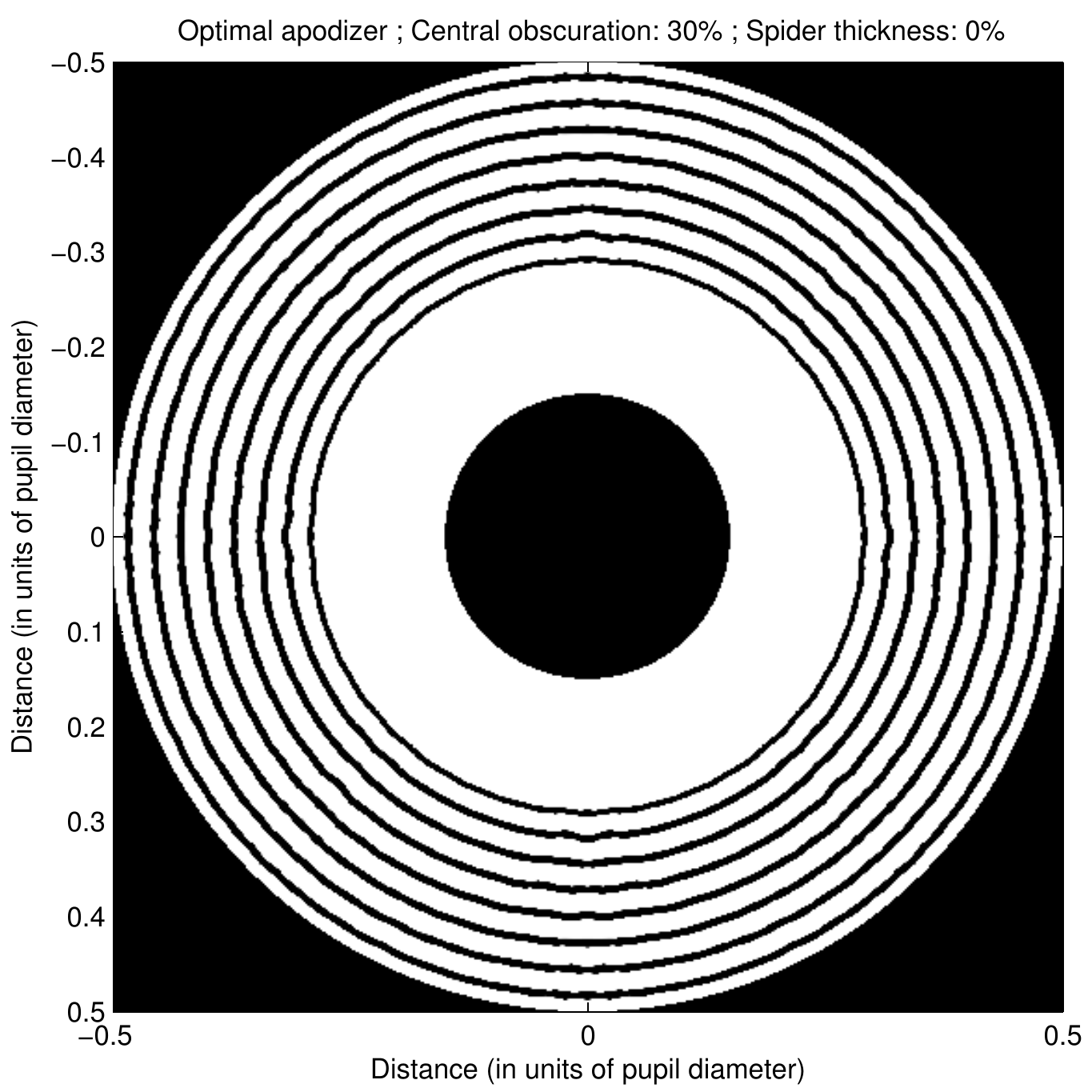}} 
\subfigure[]{\includegraphics[width=0.3\textwidth]{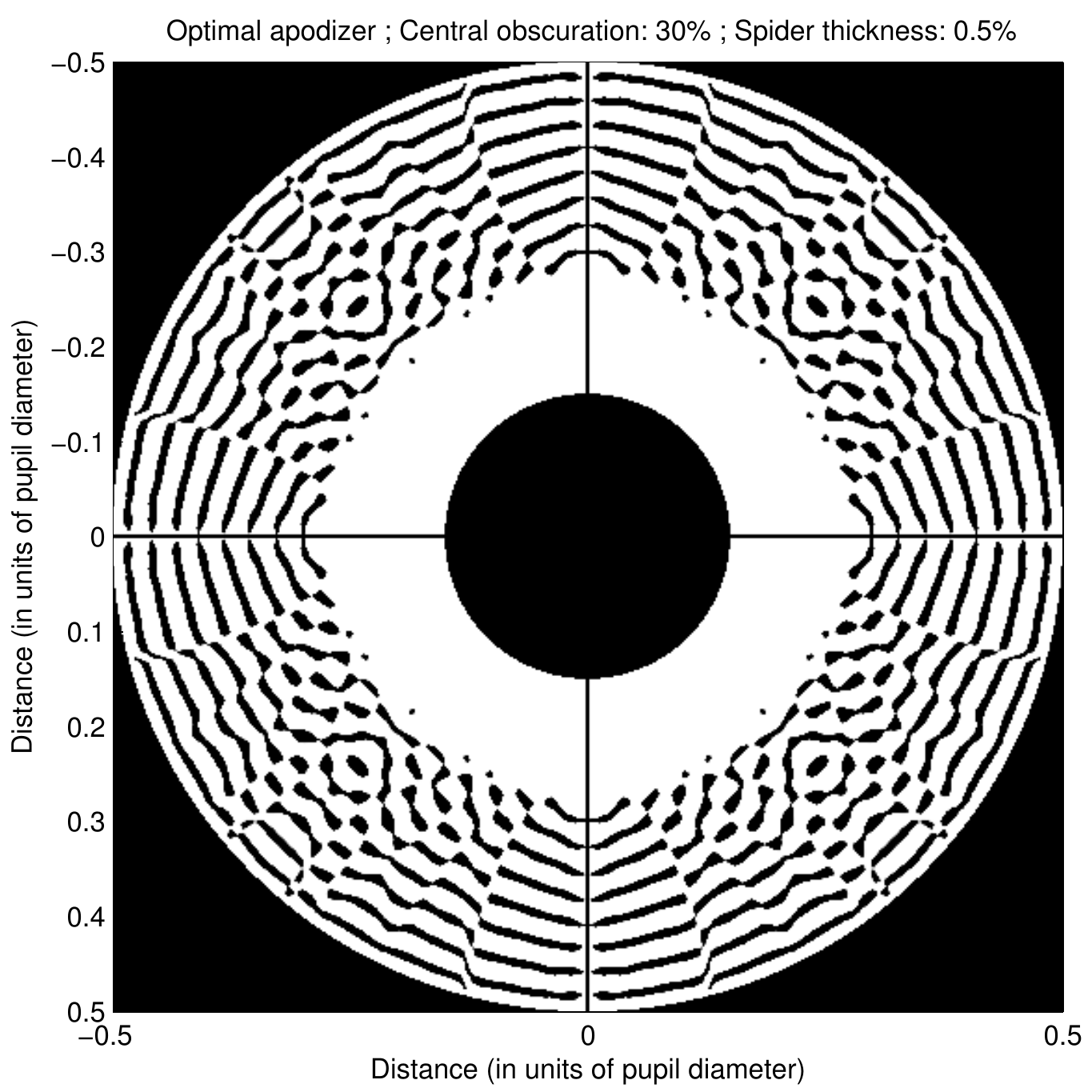}}
\subfigure[]{\includegraphics[width=0.3\textwidth]{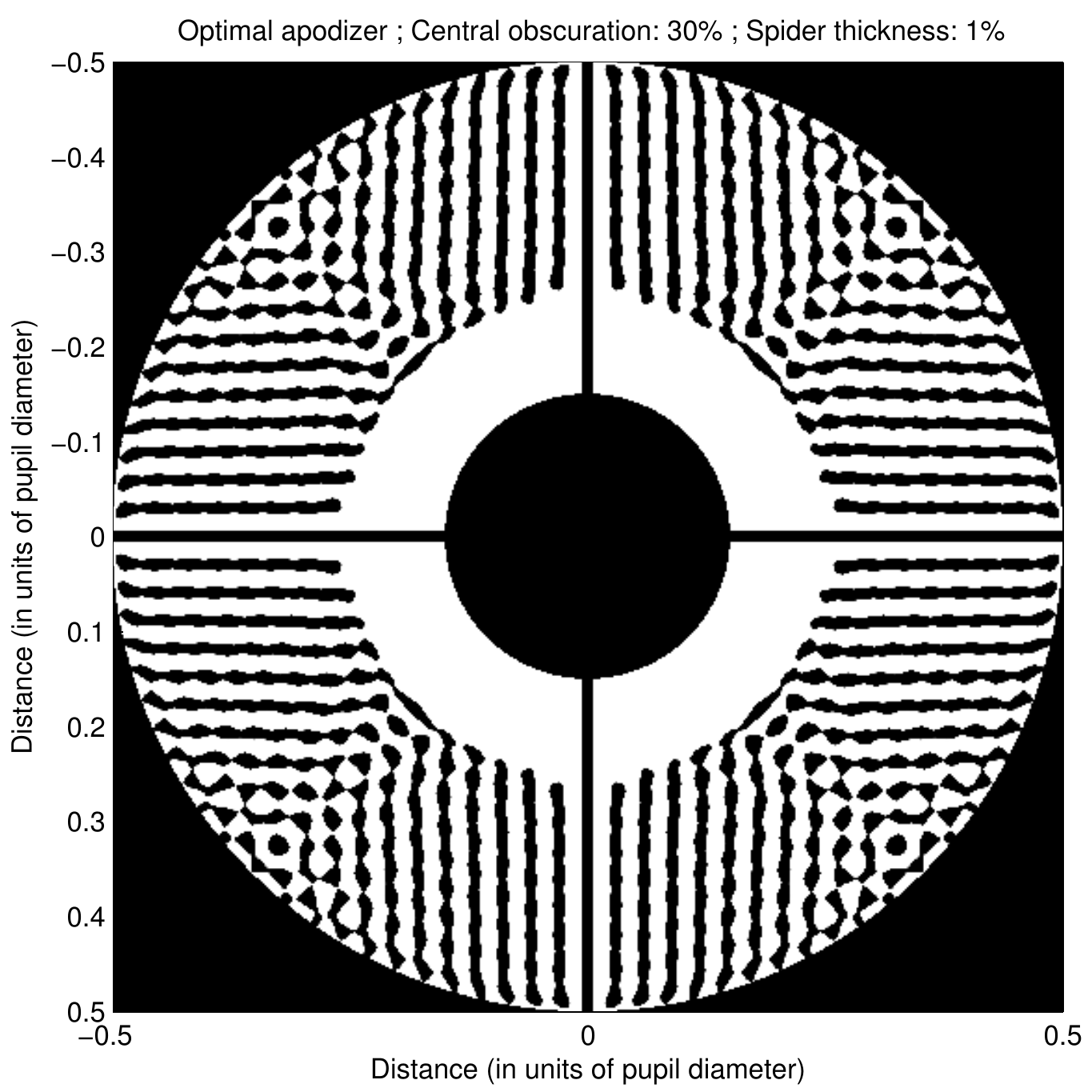}}
\end{tabular}
\caption{Apodizers giving the highest system throughput for a charge 2 VC used with a 10\% (top) and a 30\% (bottom) central obscuration, and three spider thicknesses: 0, 0.5, and 1\% (from left to right).}
\label{FigOptimalApodizersCharge2}
\centering
\end{figure}
\begin{figure}[]
\centering
\begin{tabular}{ccc}
\subfigure[]{\includegraphics[width=0.3\textwidth]{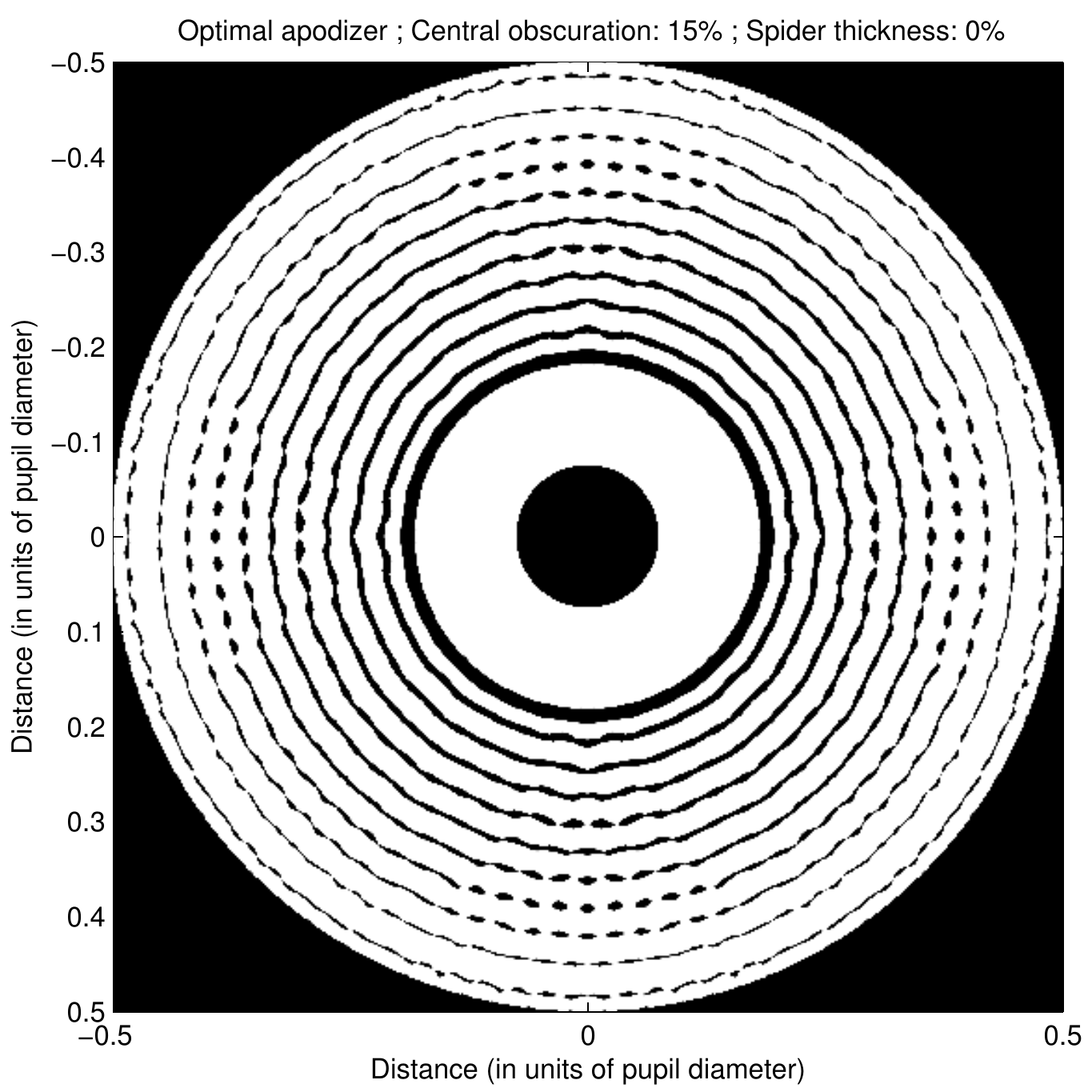}} 
\subfigure[]{\includegraphics[width=0.3\textwidth]{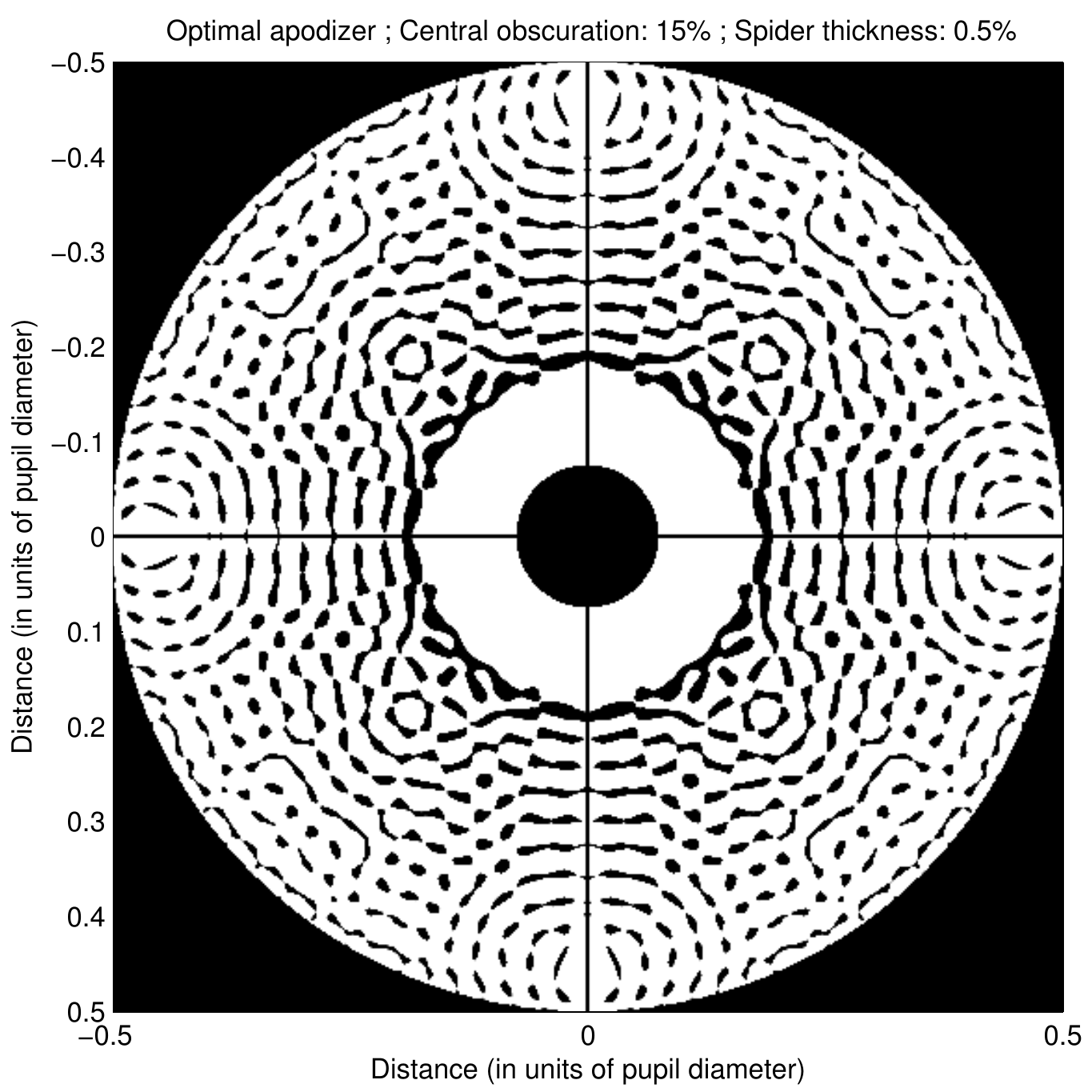}} 
\subfigure[]{\includegraphics[width=0.3\textwidth]{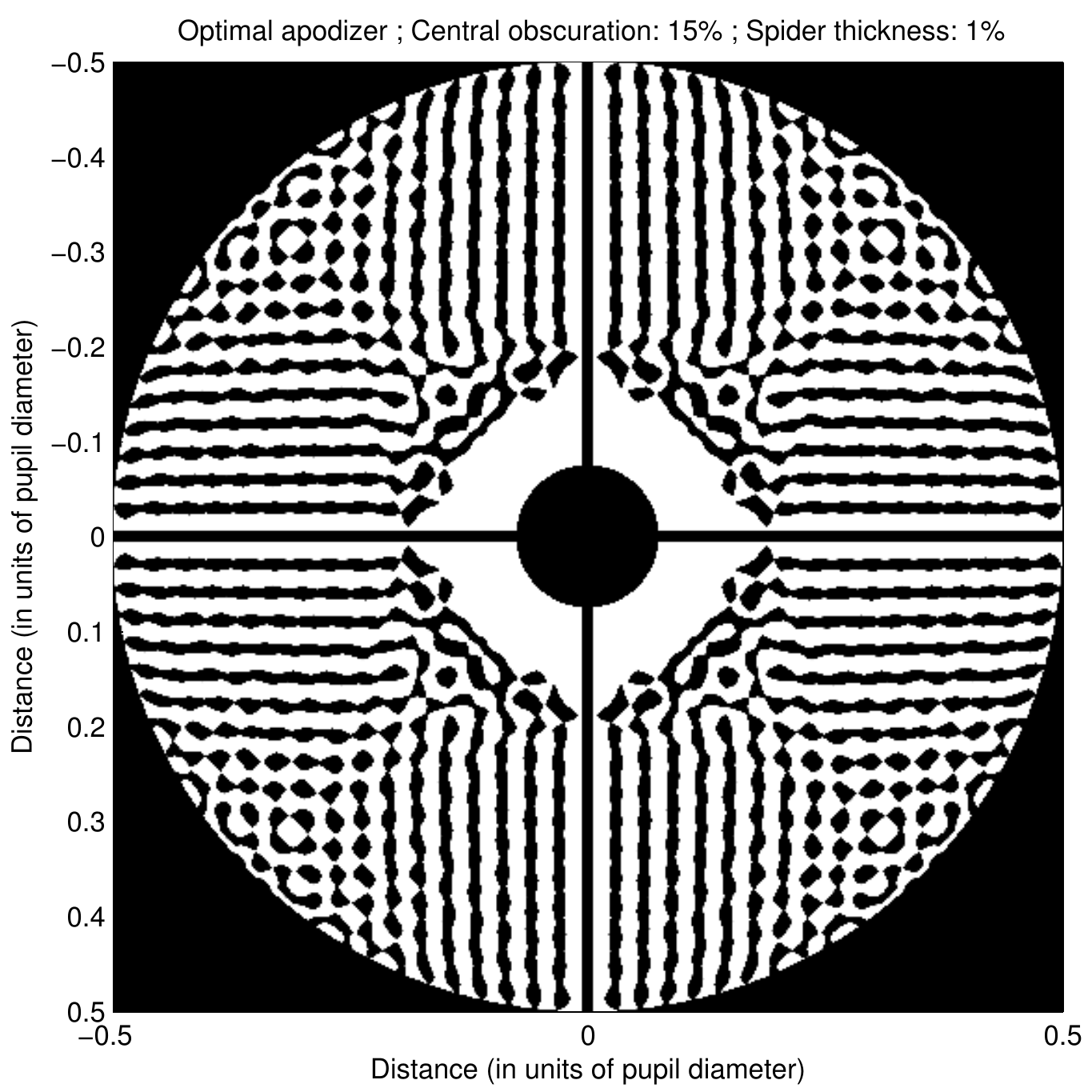}} \\
\subfigure[]{\includegraphics[width=0.3\textwidth]{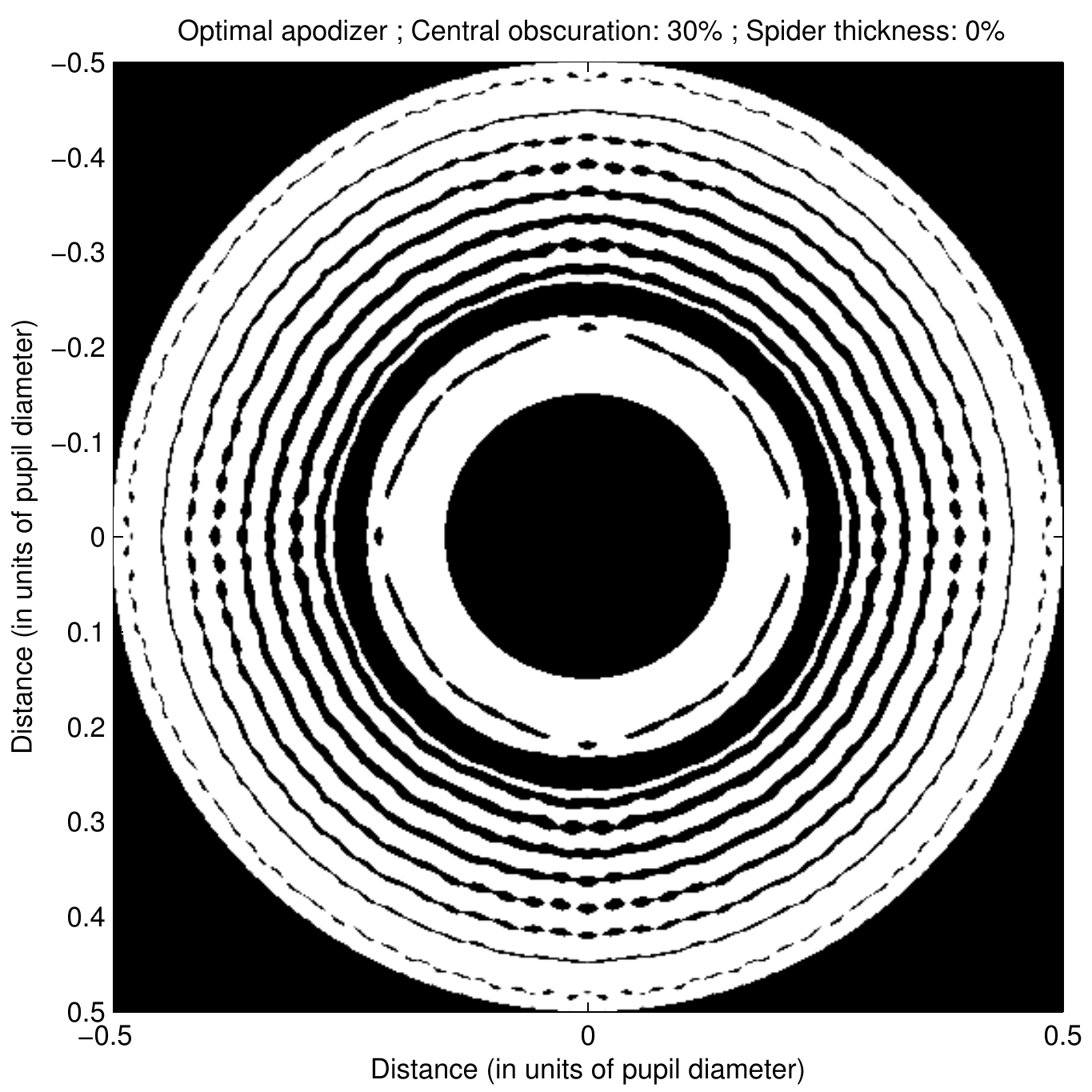}} 
\subfigure[]{\includegraphics[width=0.3\textwidth]{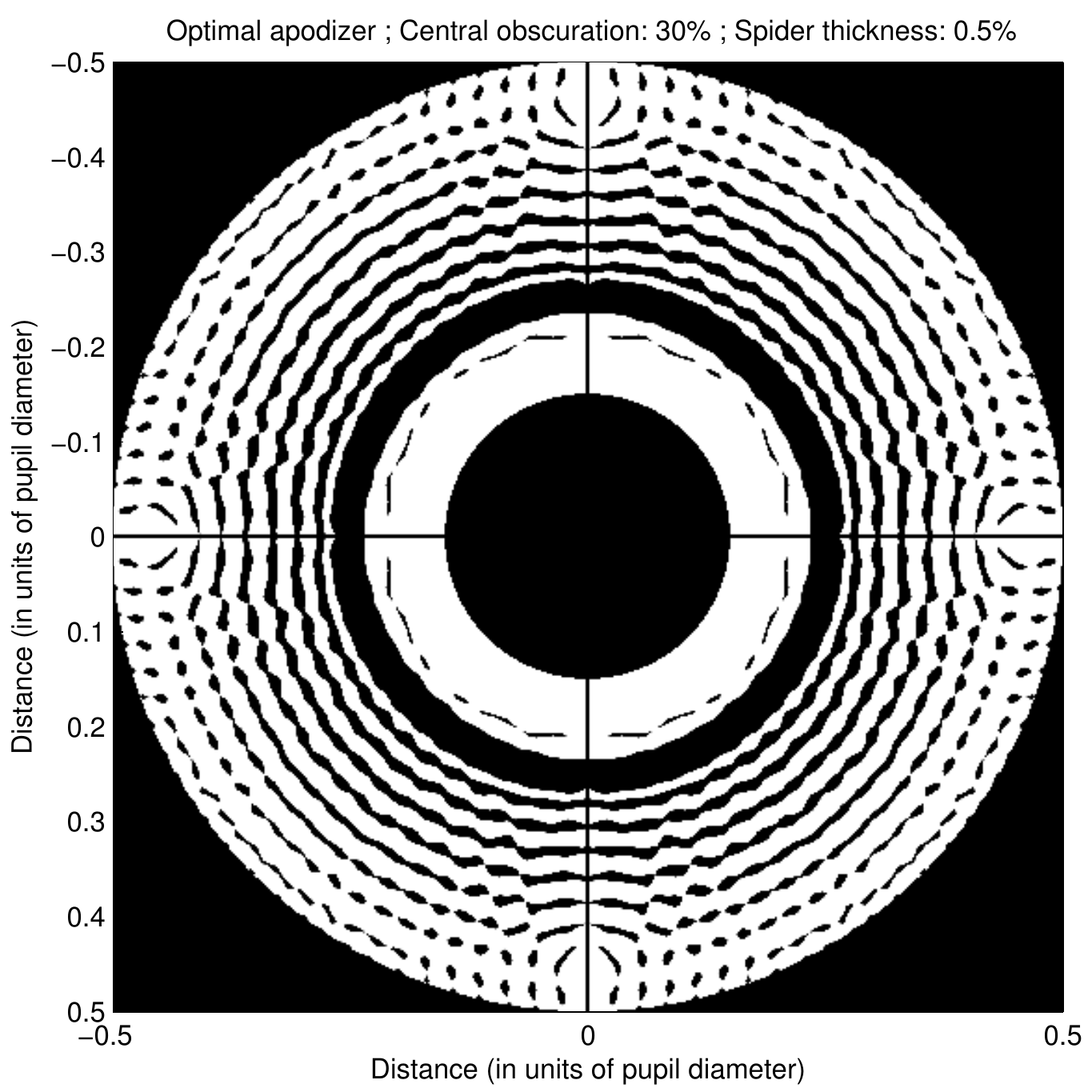}}
\subfigure[]{\includegraphics[width=0.3\textwidth]{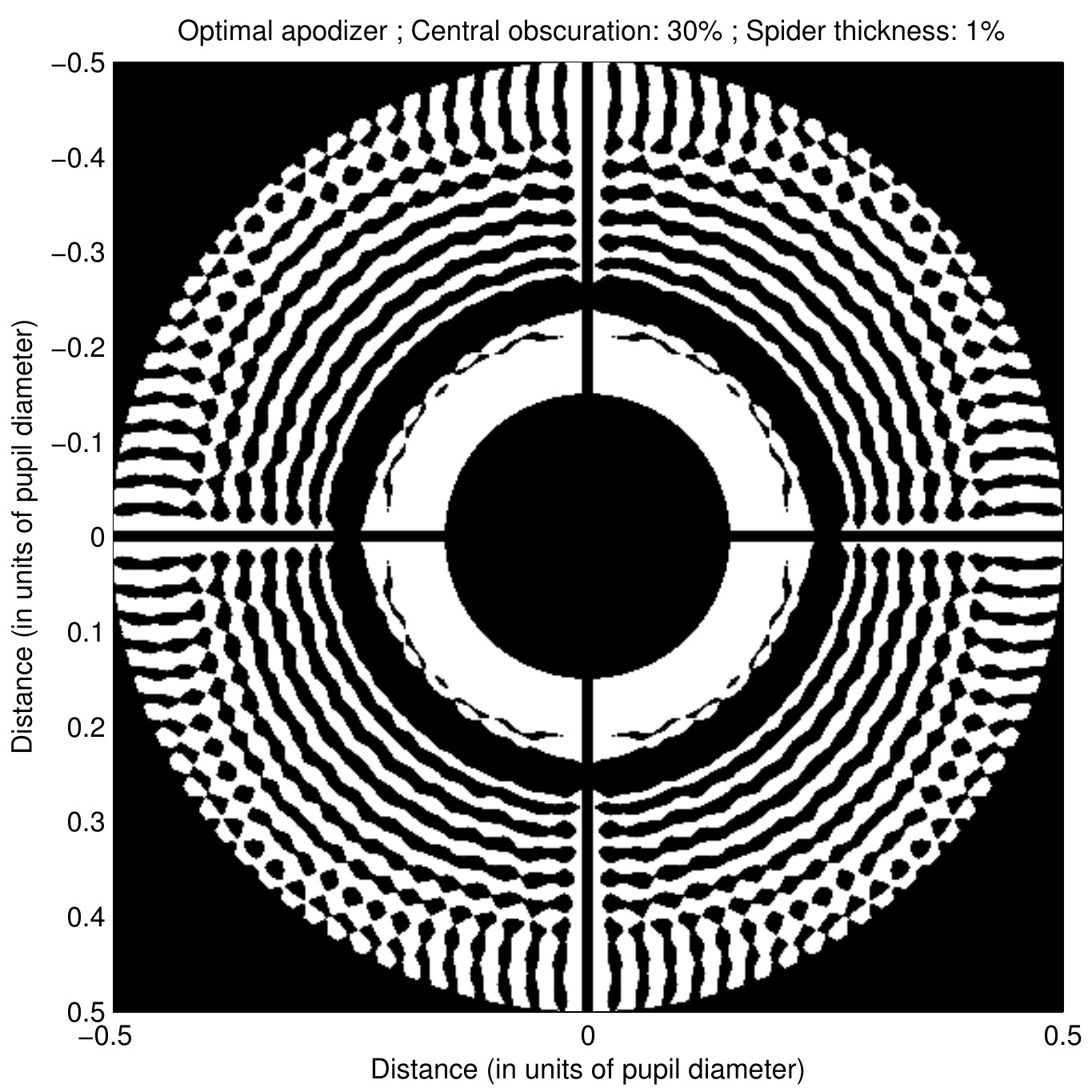}}
\end{tabular}
\caption{Apodizers giving the highest system throughput for a charge 4 VC used with a 10\% (top) and a 30\% (bottom) central obscuration, and three spider thicknesses: 0, 0.5, and 1\% (from left to right).}
\label{FigOptimalApodizers}
\centering
\end{figure}

Fig.\ref{FigOptimalApodizersCharge2} and Fig.\ref{FigOptimalApodizers} show a total of twelve apodizers optimized for charge 2 and charge 4 VC, for apertures with 10 and 30\% central obscurations, and 0, 0.5 and 1\% thick spiders. The apodizers designed for the spider-free apertures display concentric dark rings which thickness increases towards the center of the apodizer. The rings appear to be apodizing the central obscuration of the aperture.

For a charge 2 VC, one can noticed two different areas:
\begin{itemize}
\item a first annulus that surrounds the central obscuration of the aperture, where the transmission is close to 1. The outer diameter of this annulus is set by the Lyot stop for which the mask is designed.
\item a second annulus than encircles the first, where, for small central obscurations, small blocking regions are located on a 2D lattice with a 1/OWA periodicity (in units of pupil diameter). Most of these regions are isolated, but some merge into larger structures.
\end{itemize}

For a charge 4 VC, three areas are noticeable:
\begin{itemize}
\item a first annulus where the transmission is 1 almost everywhere.
\item a dark ring that surrounds the first region, which diameter is set by the Lyot stop for which the mask is designed.
\item a second annulus than encircles the two other regions. Here again, small blocking structures appear on the same 1/OWA periodic 2D lattice.
\end{itemize}

The presence of spiders creates a more complicated pattern, as the mask apodize both the central obscuration and the spiders. It is interesting to compare the patterns displayed by the apodizers designed for a charge 4 VC, and for apertures with 1\% thick spiders and 10 and 30\% central obscurations: in the case of the smaller obscuration, thick horizontal and vertical dark strips can be seen next to the spiders, i.e., the apodizer appears to be mainly apodizing the spiders. On the other hand, when looking at the other apodizer, thick dark concentric rings can be seen, and the apodizer appears to be mainly apodizing the central obscuration.

\subsection{Inner working angle} \label{OFFAXISPERF}

The main advantage of phase mask coronagraphs is their very small IWA. As noted in previous papers, using an apodizer changes the usual off-axis transmission of the coronagraph.

The maximum throughput $T_{max}$ is reached for sources at around 25-26 \ld from the star, but $T_{max}/2$ is reached at much shorter distances ($\approx$1-3\ld depending on the topological charge, the central obscuration, and the spider thickness of the aperture). This half-throughput is usually used to define the IWA of the coronagraph (as suggested in \cite{Guyon2006}). 

\begin{figure}[]
\centering
\begin{tabular}{c}
\subfigure[]{\includegraphics[width=0.3\textwidth]{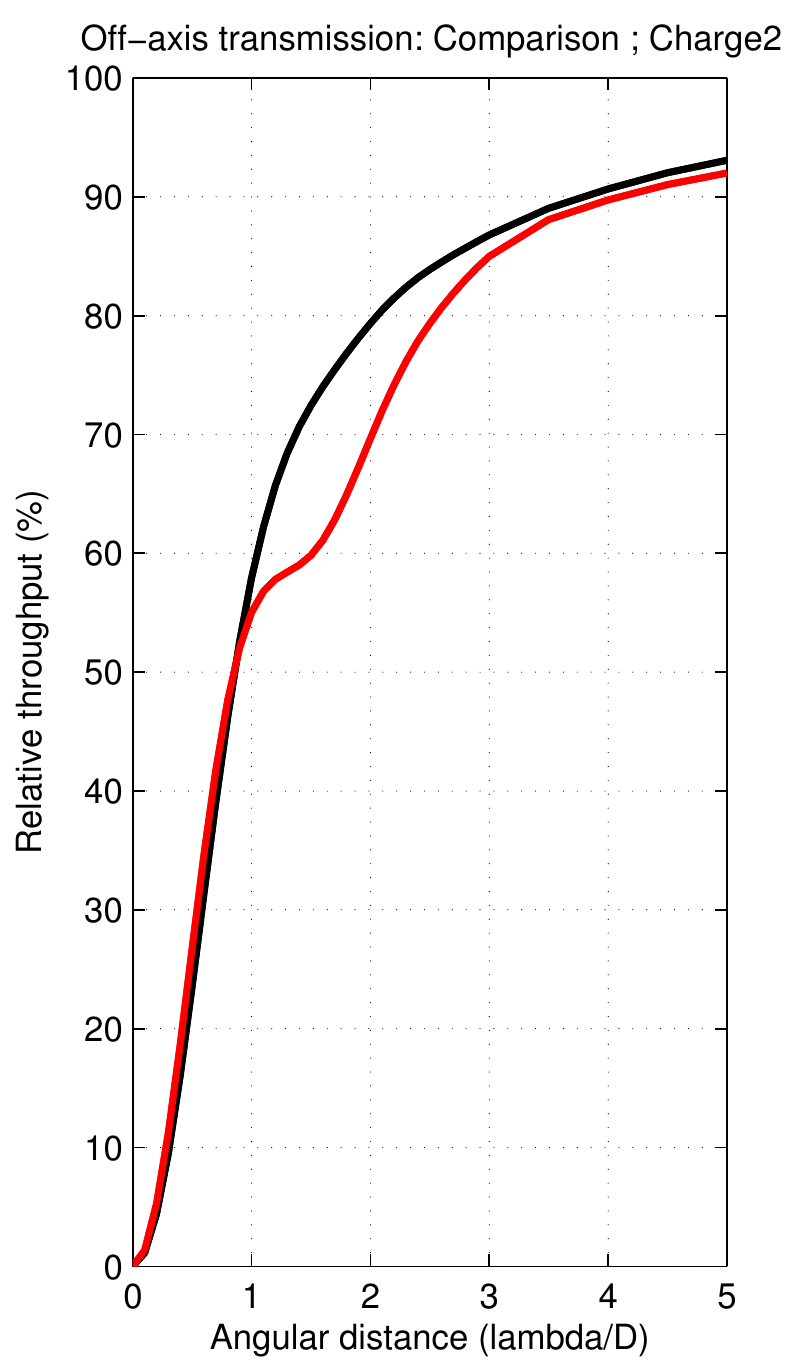}}
\subfigure[]{\includegraphics[width=0.3\textwidth]{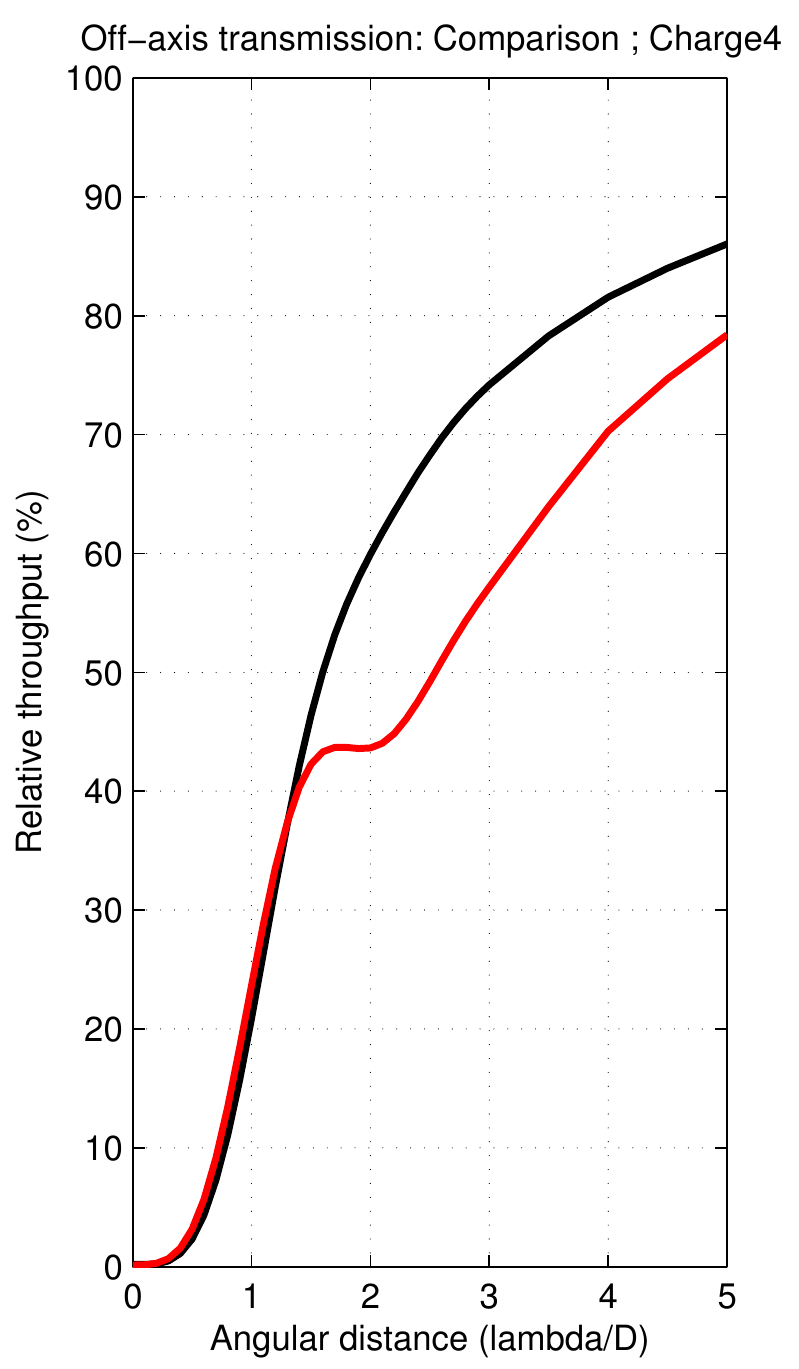}}
\end{tabular}
\caption{Comparison of the relative off-axis transmission of the VC (black line) and an AVC (red line) designed for a $CO_{A}=20\%$ obscured aperture with $t_{S}=0.5\%$ spiders. The comparison is made for (a) a charge 2 vortex and (b) a charge 4 vortex. The transmissions are normalized with respect to the coronagraph throughput $T_{max}$.}
\label{TransmissionComparison}
\centering
\end{figure}

A first and important observation must be made: as illustrated in Fig.\ref{TransmissionComparison}, and contrary to the off-axis transmission curve of a VC, the transmission of an AVC is not monotonic. Because of that, in some cases, using the IWA to fully describe the resolution properties of the AVC may be difficult. 

For angular distances $\theta$ between 0 and 1-1.2\ld, the transmission (normalized with respect to the maximum coronagraphic throughput $T_{max}$) follows a $\theta^{l}$ function, with $l$ the topological charge of the VC. This confirms that, like the VC, the AVC is a second or fourth-order coronagraph.

In Fig.\ref{TransmissionComparison}, the apodizer of the AVC is designed for a 20\% obscured aperture with 0.5\% thick spiders. The transmission increase rate slows down between 1 and 1.6\ld for $l=2$, and between 1.5 and 2.2\ld for $l=4$. This does not have much effect on the IWA of the charge 2 AVC: it is almost identical to what it is for the charge 2 VC (0.9\ld). The same is not true for the charge 4 AVC, however: while 40\% of the maximum throughput is reached for $\theta=1.4\ld$, the IWA of this coronagraph is about 2.5\ld (while it is 1.75\ld with a charge 4 VC).

\begin{figure*}[]
\centering
\begin{tabular}{c}
\subfigure[]{\includegraphics[width=0.3\textwidth]{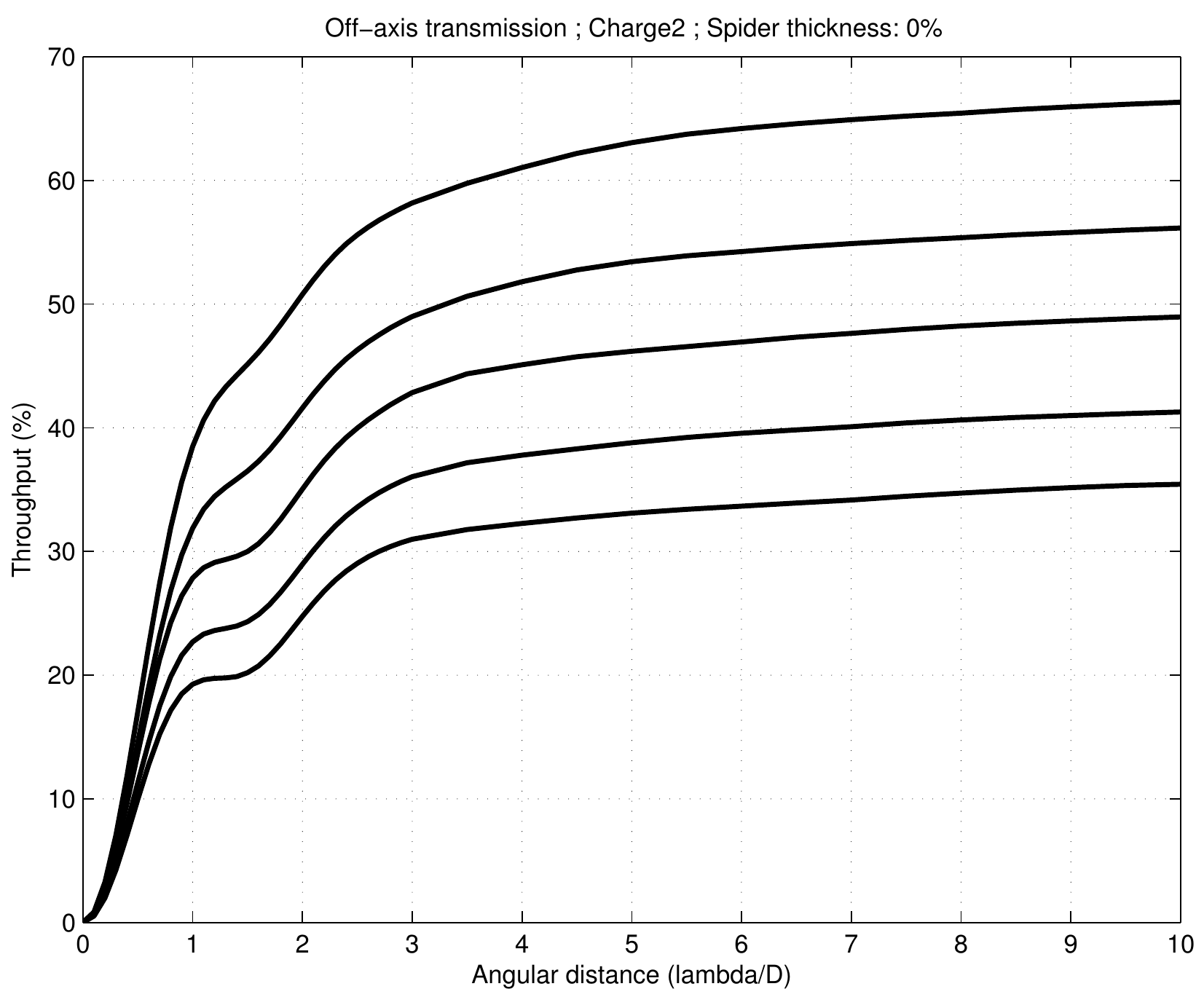}}
\subfigure[]{\includegraphics[width=0.3\textwidth]{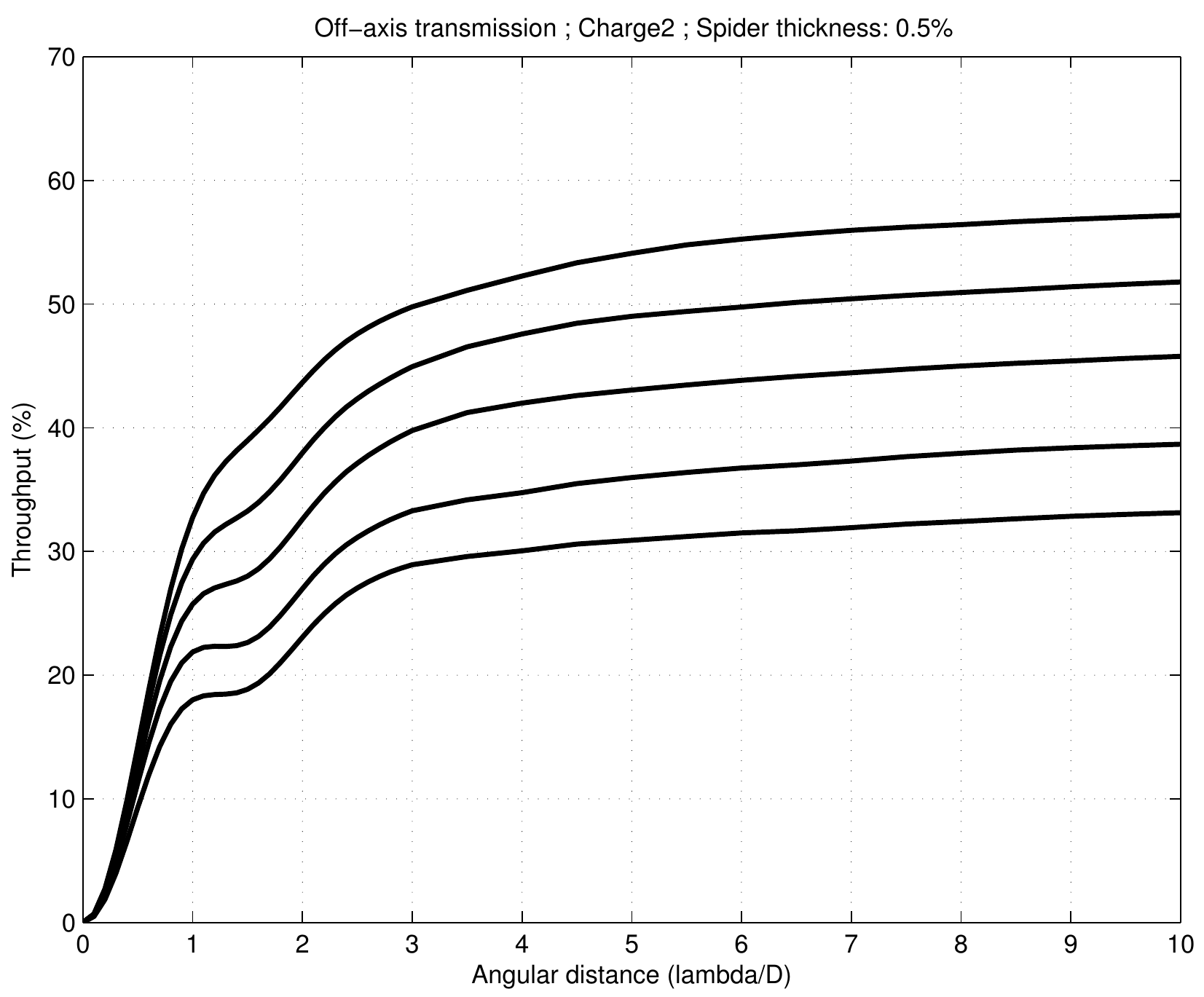}}
\subfigure[]{\includegraphics[width=0.3\textwidth]{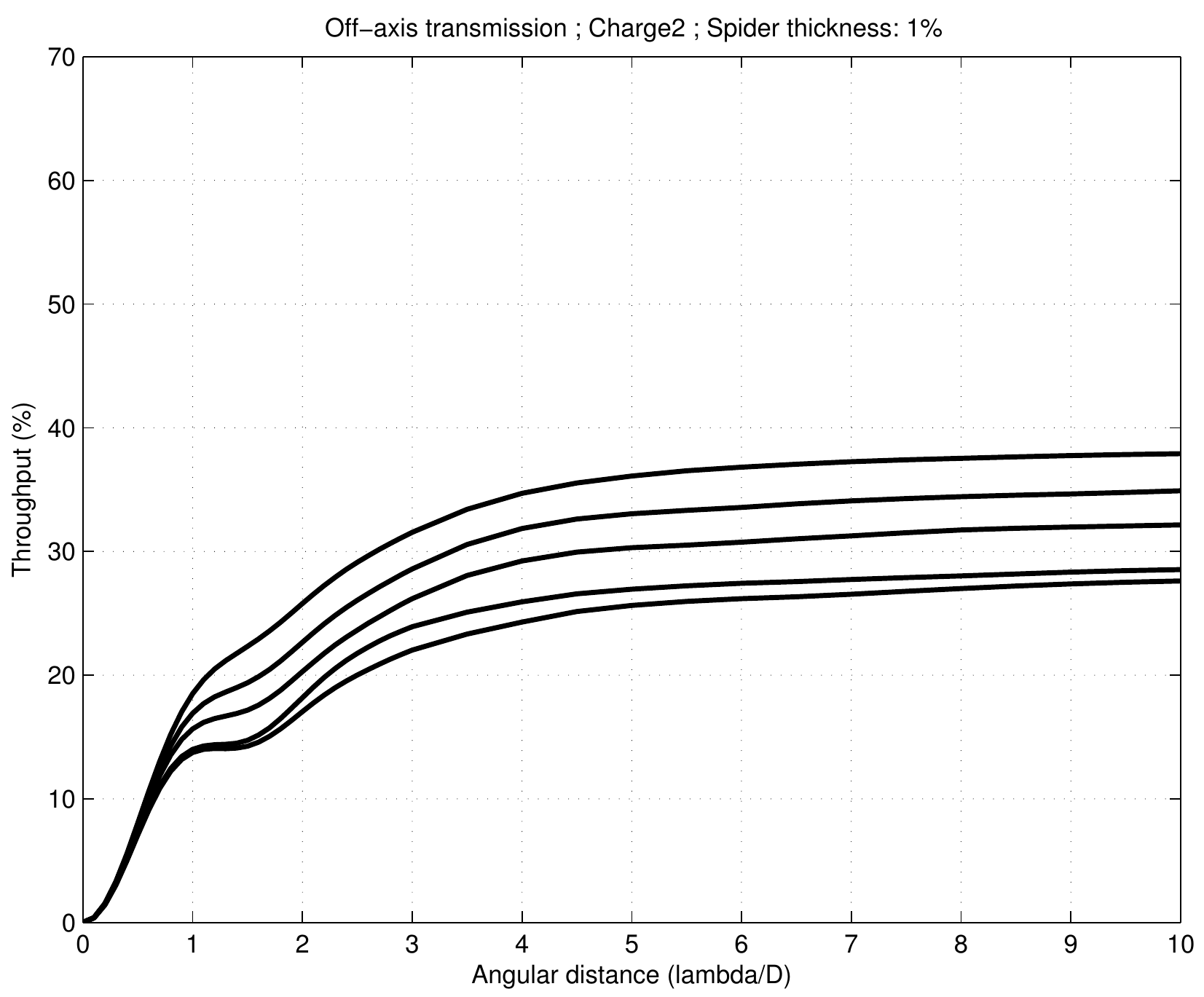}}\\
\subfigure[]{\includegraphics[width=0.3\textwidth]{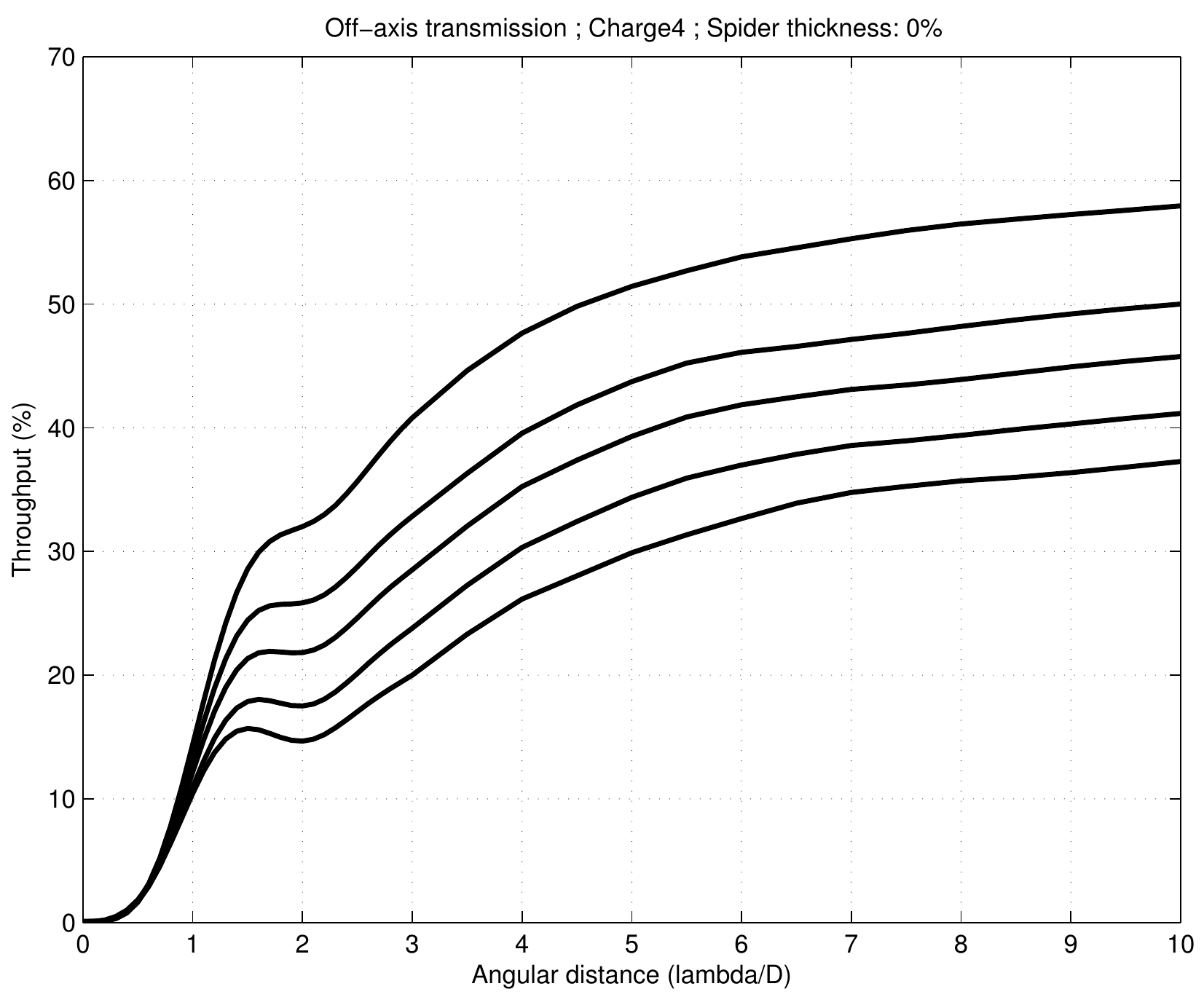}}
\subfigure[]{\includegraphics[width=0.3\textwidth]{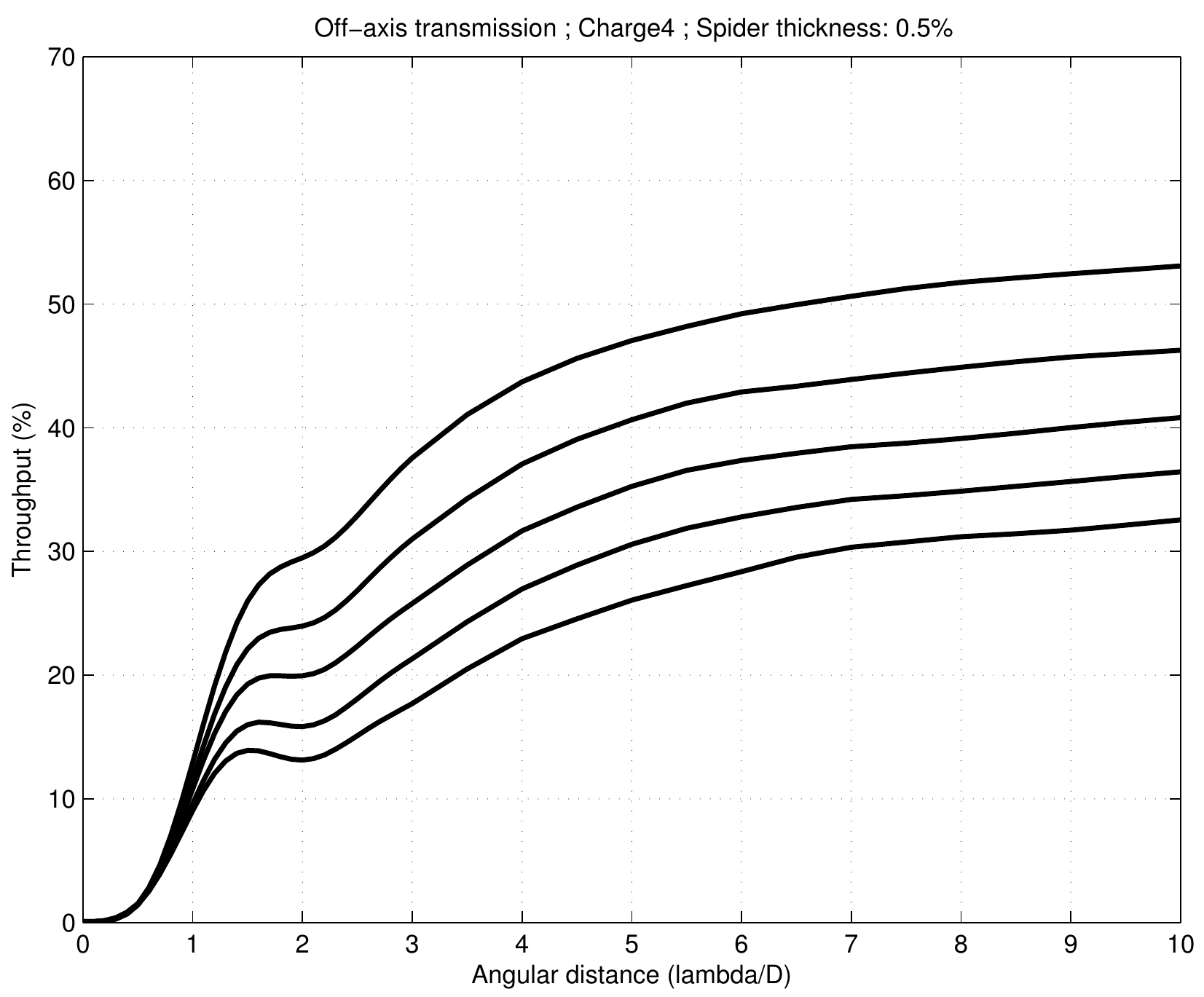}}
\subfigure[]{\includegraphics[width=0.3\textwidth]{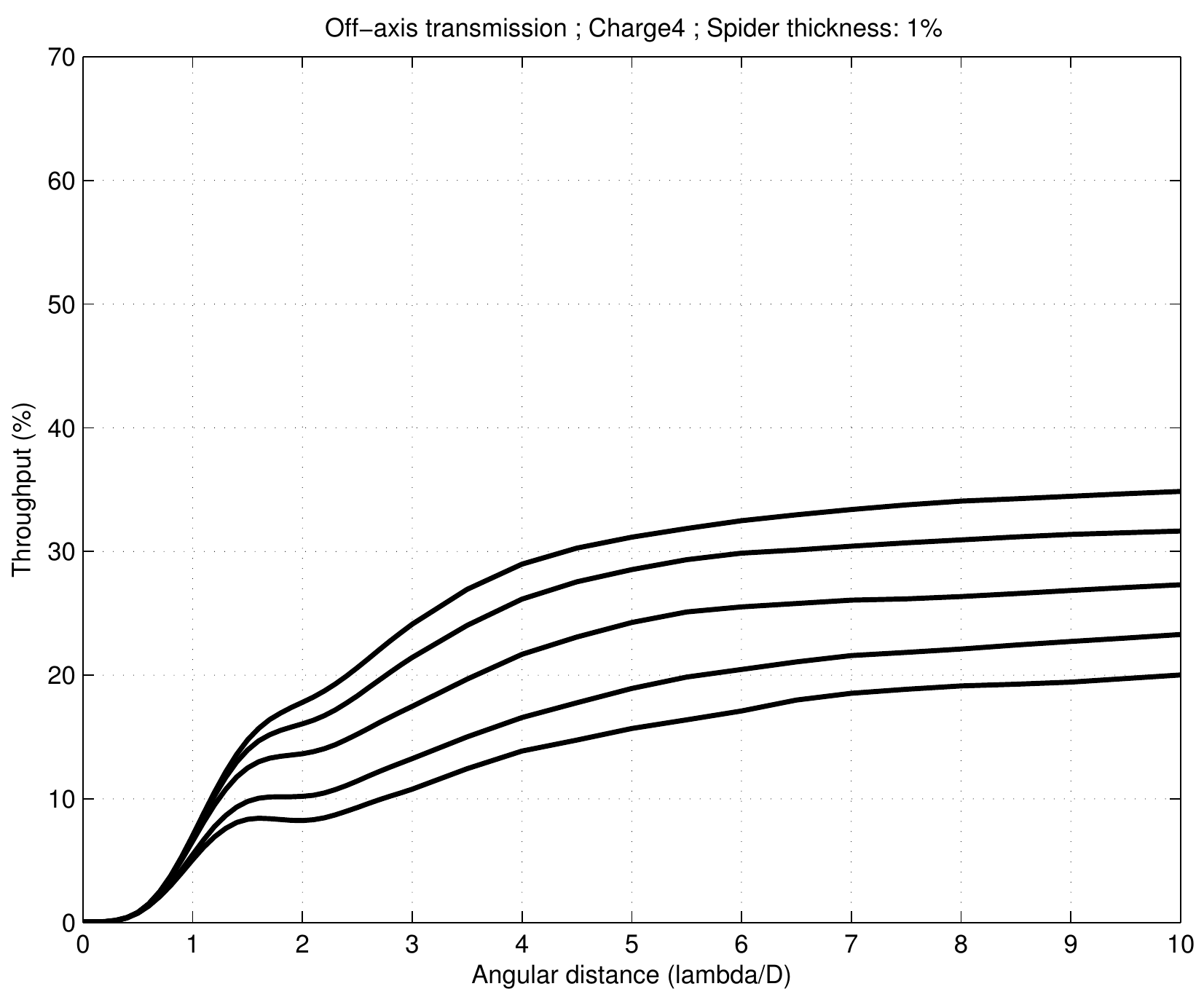}}
\end{tabular}
\caption{Transmission (in \%) of an off-axis source as a function of the distance to the star (in units of \ld), for (a) \& (d) apertures with no spiders, (b) \& (e) apertures with 0.5\% thick spiders, and (c) \& (f) apertures with 1\% thick spiders. Figures (a), (b), and (c) correspond to a charge 2 VC, and figures (d), (e), and (f) correspond to a charge 4 VC. In each case, the transmission is given for the five central obscurations: 10, 15, 20, 25, and 30\% (from top to bottom). The red dots indicate the IWA of each coronagraph.}
\label{TransmissionCurves}
\centering
\end{figure*}

Fig.\ref{TransmissionCurves} shows how the off-axis transmission for the five different central obscurations and the three different spider thicknesses that were considered. Note that these numbers are also reported in Tab. \ref{tabZERO} and \ref{tabONE}. Again, the transmission slows down at about 1-1.2\ld, before increasing again at about 1.5-2\ld. The effect strengthens with the obscuration ratio and the topological charge. For $l=2$, the transmission slows down for $CO_{A}$=10\%, and it stalls for $CO_{A}$=30\%. For $l=4$, it stalls for $CO_{A}$=10\%, and it decreases for $CO_{A}$=30\%.

For $l=2$ the IWA of the AVC is close to that of the VC for $t_{S}$=0\% and 0.5\%, where the half-transmission is reached for $\theta$=0.9\ld. The IWA only differs for $t_{S}$=1\% for which it increases with the obscuration, ranging between 1.1 and 1.5\ld. For $l=4$ the IWA of the AVC is larger than what it is with the VC. The half-transmission is reached for $\theta$=1.8-3.2\ld for $t_{S}$=0\% and 0.5\%, and for $CO_{A}$=10-30\%. For $t_{S}$=1\%, it ranges between 2.2 and 3.2\ld.


\begin{figure}[]
\centering
\includegraphics[width=0.55\textwidth]{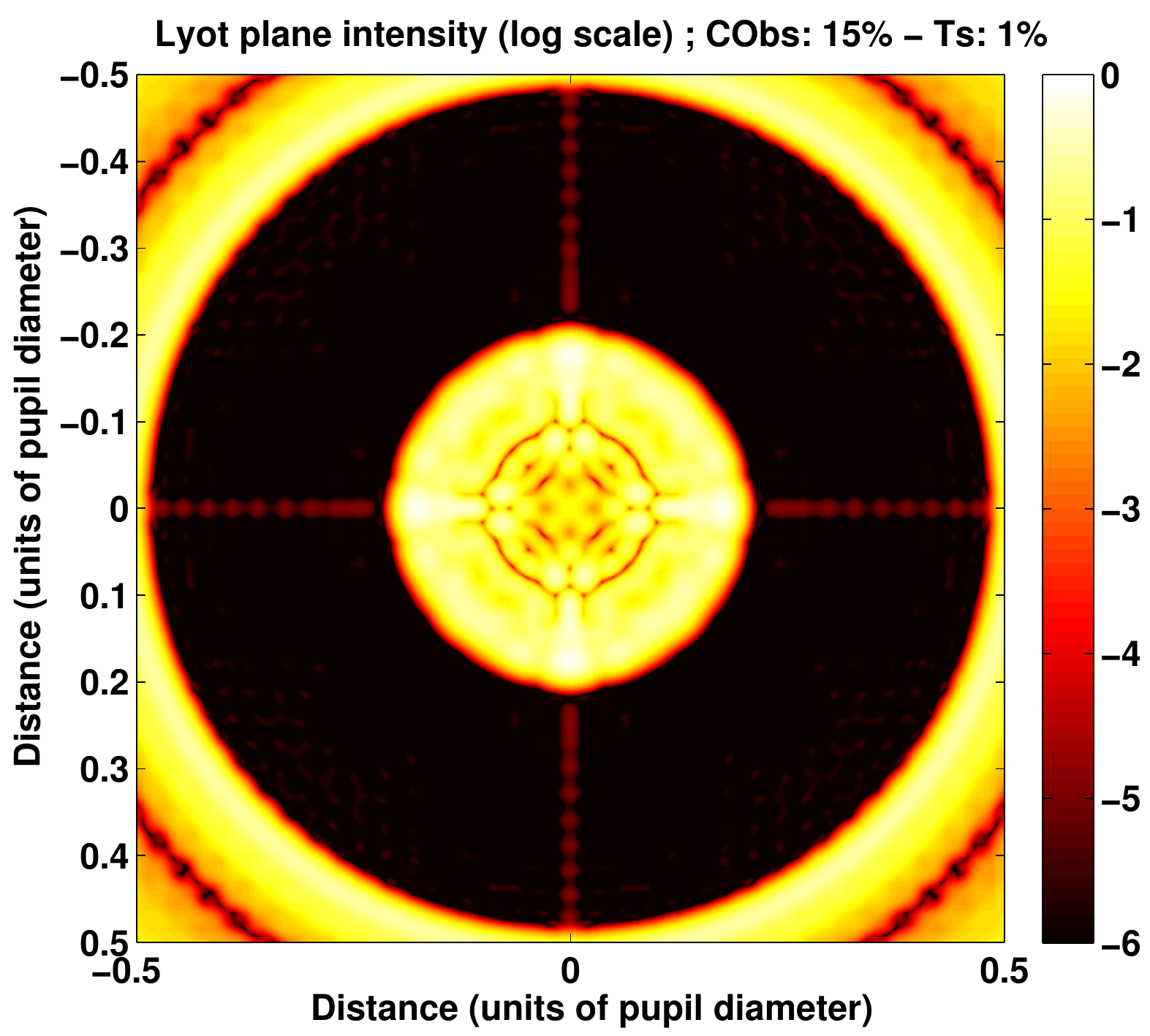}
\caption{Example of a 2D intensity distribution in the Lyot plane of the apodized VC. The intensity is displayed on a log scale for an aperture with a 15\% central obscuration and 1\% thick spiders.}
\label{LyotPlane}
\centering
\end{figure}

Associated with the apodizer, the vector vortex masks attenuate the on-axis light of the star in a given region of the Lyot plane. Figure \ref{LyotPlane} illustrates one of the 2D intensity distributions that have been computed. In the dark zones that appears in the Lyot plane, the intensity has been reduced to a few millionth of what it would have been without the apodizer.

\begin{figure}[]
\centering
\includegraphics[width=0.55\textwidth]{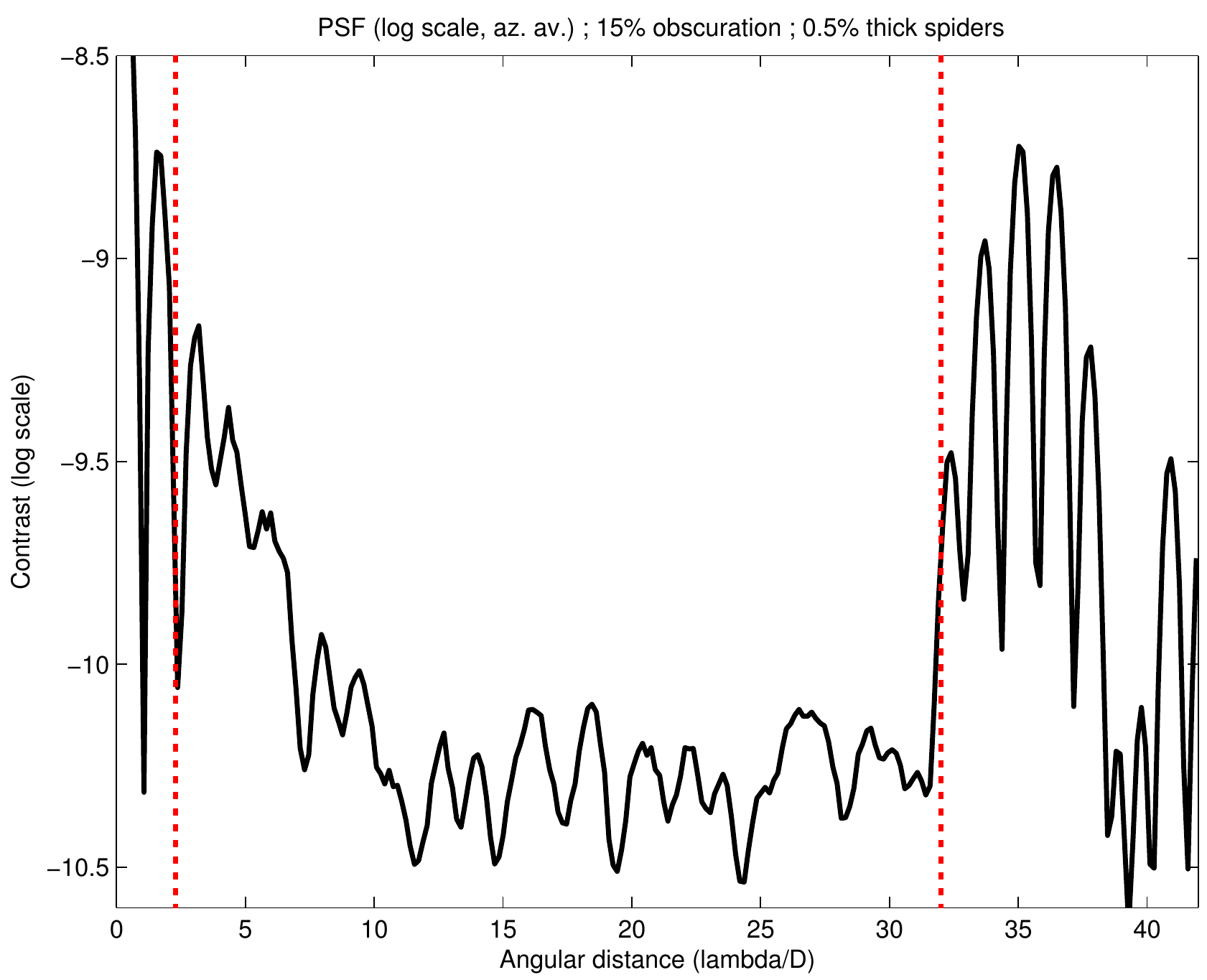}
\caption{Coronagraphic PSF (azymutal average) for the apodized vortex designed for a 15\% centrally obscured aperture and 0.5\% thick spiders. The intensity is displayed on a log scale. Angular distances are given in units of \ld.}
\label{PSF}
\centering
\end{figure}

The point-spread function (PSF) of the coronagraph is computed by taking the Fourier transform of the product of the appropriate Lyot stop and the electric field observed in the Lyot plane. Normalizing this PSF is done by computing the propagation of an off-axis source (located as close to the OWA as it is possible without being attenuated). The intensity of the first PSF is divided by the maximum intensity of the second one. An example of such a PSF is showed in Fig. \ref{PSF}. One can see that a central structure extends up to 12-15 \ld from the star, where the intensity is lower than $3 \times 10^{-9}$. Beyond this distance, the intensity stays close to $3 \times 10^{-10}$ (there is no point in looking further away than 32 \ld since that distance is also the radius of the vortex phase mask).

Tab. \ref{tabZERO} and \ref{tabONE} list the imaging properties of the 15 coronagraphs, and in particular the contrast measured at the IWA (it is an average value computed in an annulus with a mean radius that equals the IWA, and a width of 1\ld). The contrast in the image plane is only indirectly constrained: the real constraints is set on the intensity of the electric field in the Lyot plane. 

The contrast is not as low with the $l=2$ AVC than it is with the $l=4$ AVC. For $l=2$ it is between $5\C{9}$ and $8\C{8}$, and neither the obscuration ratio nor the spiders thickness appear to have a strong influence on the contrast (its mean value is 5\C{8} for the spider-free apertures, 1.5\C{8} for the $t_{S}=0.5\%$ apertures, and 5\C{8} for the $t_{S}=0.5\%$ apertures). 

It can be noticed that the contrast increases with both $CO_{A}$ and $t_{S}$. The contrast is close to 2-3$\times 10^{-10}$ for the spider-free apertures, except for $CO_{LS}=30\%$ where it is 6$\times 10^{-10}$. It becomes about 10 times higher for $t_{S}=0.5\%$, with a maximum of 7$\times 10^{-9}$. Finally, for $t_{S}=1\%$, the maximum contrast becomes 2-3$\times 10^{-8}$.

The contrast increase is partially explained by the fact that in the optimizations the intensity measured in the Lyot plane is not normalized by the transmission of the apodizer $T_{A}$. Nonetheless, $T_{A}$ and $T_{max}$) do not change enough from one aperture to the other to explain the contrast increase, and the constraint set on the intensity should be adjusted to improve the contrast.

\subsection{Chromaticity}\label{Chromaticity}

The most recent iterations of the vortex phase mask are partially achromatic: $10^{-8}-10^{-9}$ contrast have been measured in a 10\% band \citep{Mawet2012}. It is important to study the chromaticity of the apodized coronagraph as well.

The apodizer is designed for a focal plane mask with a diameter which is fixed when measured in units of \ld. Hence, the physical size of the mask should scale with the wavelength, while in fact it stays the same. At shorter wavelengths, the mask will appear to cover a larger area that expected, and the opposite effect will occur at longer wavelengths.

\begin{figure}[]
\centering
\begin{tabular}{c}
\subfigure[]{\includegraphics[width=0.45\textwidth]{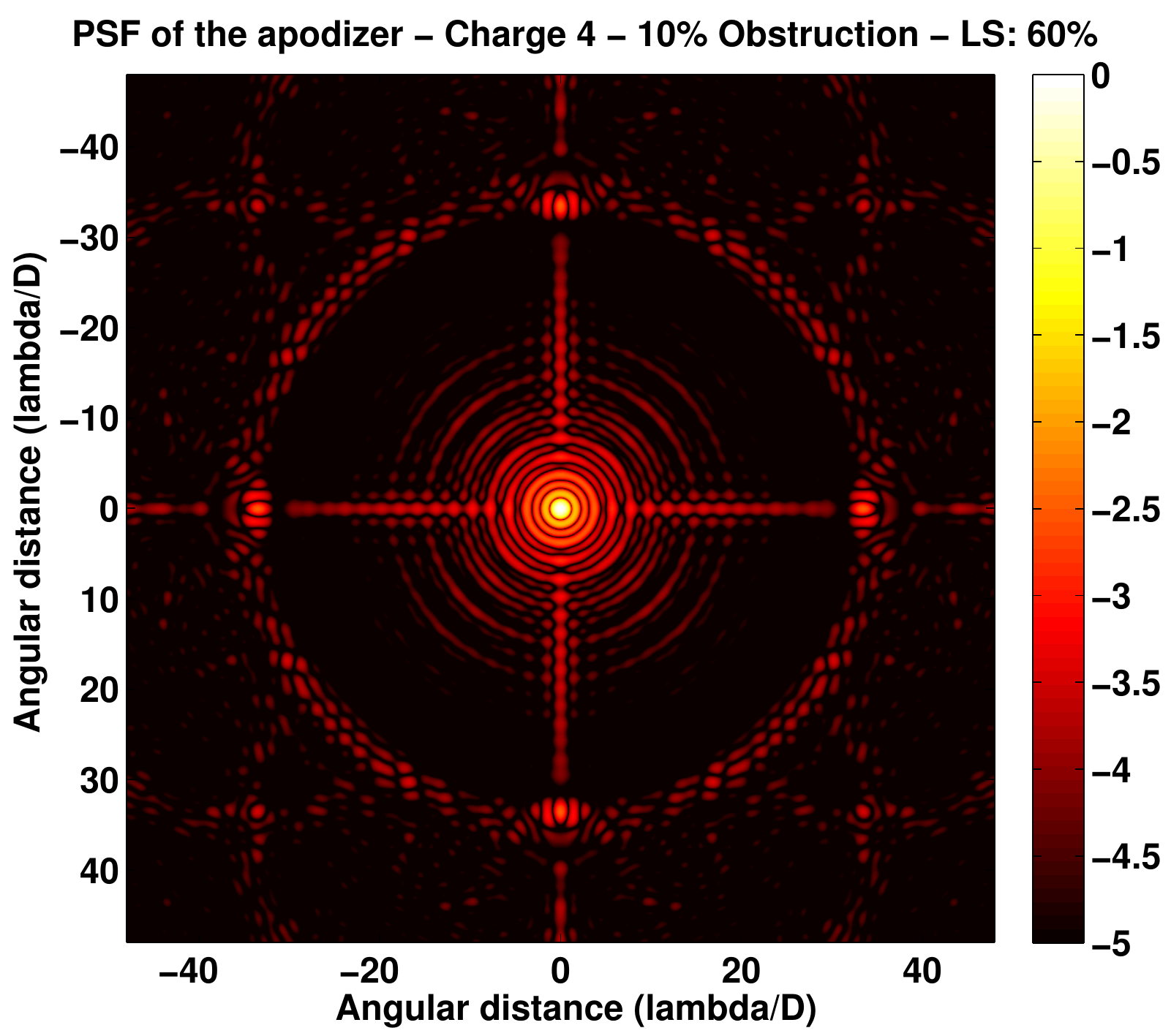}} 
\subfigure[]{\includegraphics[width=0.5\textwidth]{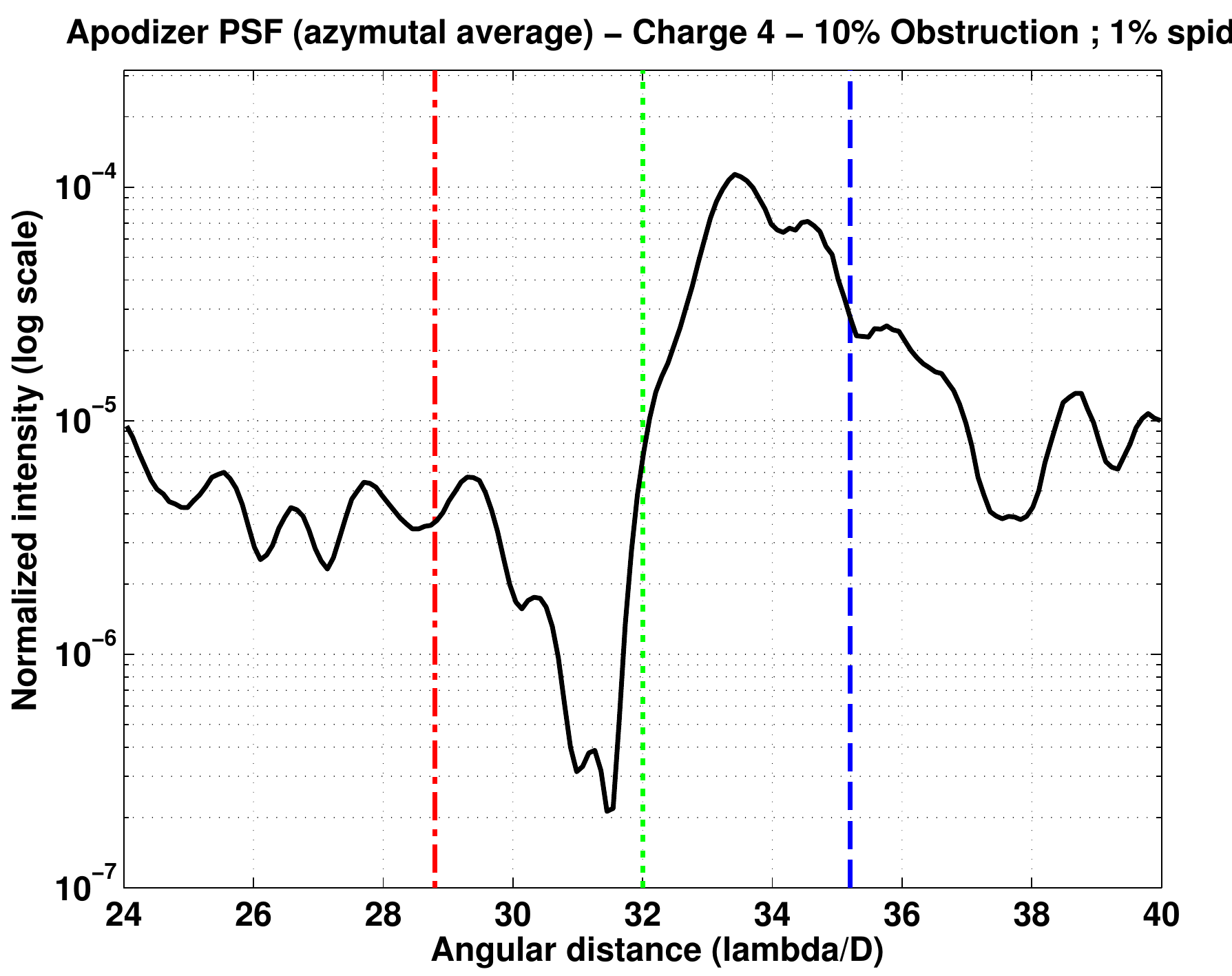}}
\end{tabular}
\caption{2D (a) and azymutal average (b) of the PSF of the optimal apodizer designed for a 10\% central obscuration, 1\% thick spiders, and a charge 4 vortex. In the azymutal average, the red dot-dashed line denotes the highest spatial frequency captured by the mask at $\lambda=1.1\lambda_{0}$, the blue dashed line does the same $\lambda=0.9\lambda_{0}$, and the dotted green line does it at $\lambda=\lambda_{0}$.}
\label{PSFchromatism}
\centering
\end{figure}

As it can be seen in Fig. \ref{PSFchromatism} a relatively bright ring surrounds the PSF of the optimal apodizer. This is quite understandable since the apodizer modifies the PSF to make it compatible with the vortex phase mask, which has a finite radius that in this case equals 32\ld. Beyond that angular distance the amplitude of the electric field has no reason to match the requirements of the vortex phase mask. The impact on the performance of the coronagraph is thus expected to be stronger at shorter than at longer wavelengths. 

It is also expected that chromatic effects may be weaker for larger vortex phase masks, as the intensity of the ring that surrounds the region for which the mask is designed will get weaker too. The current limitation of the mask radius is only due to the limited amount of RAM in the computer we used to perform the optimization (64GB). Much larger amounts are available in other machines, however, and larger masks could be computed. One should keep in mind that smaller mask radii may be responsible for higher coronagraphic throughputs, however.

\begin{figure}
\centering
\includegraphics[width=0.65\textwidth]{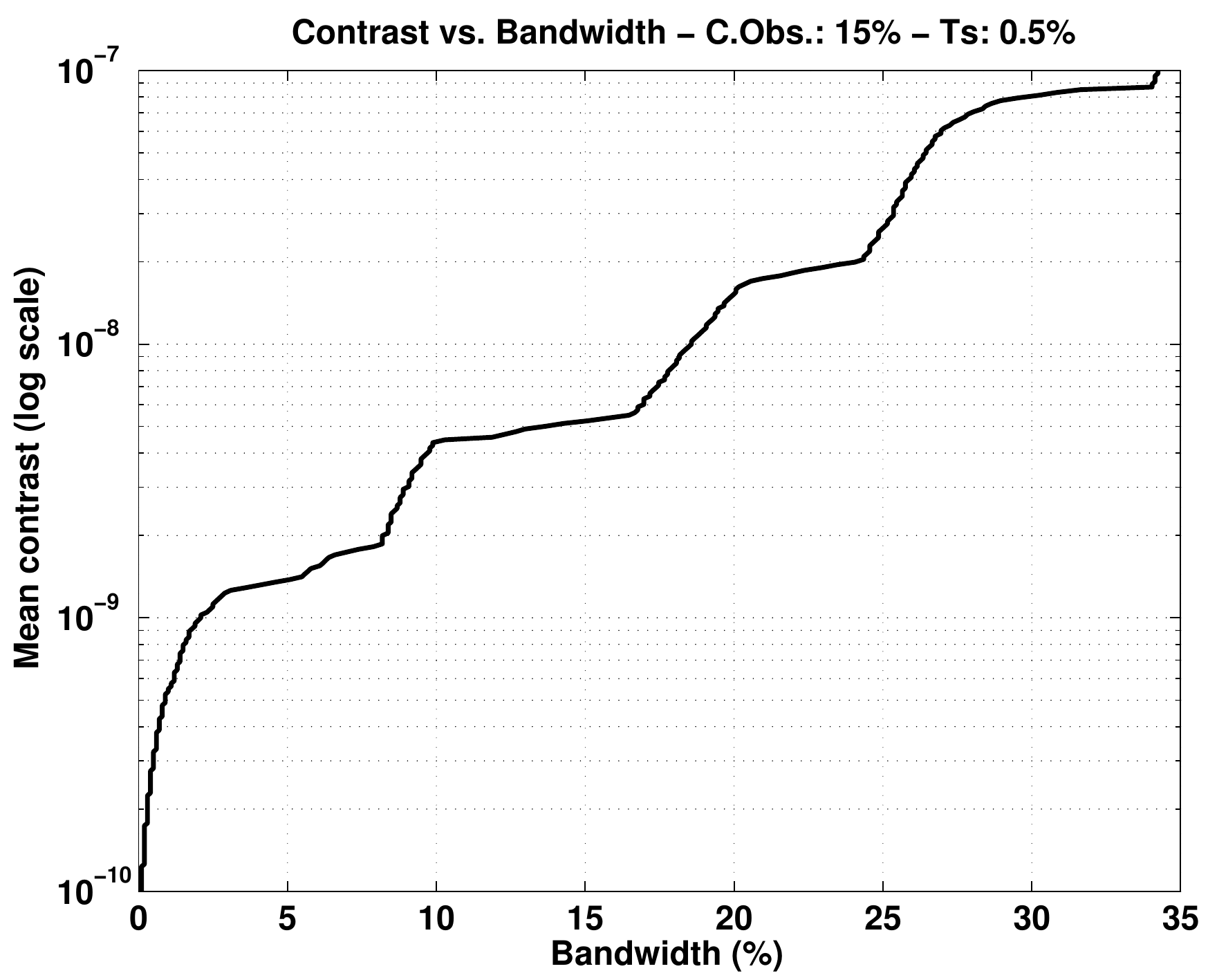}
\caption{Mean contrast as a function of the bandwidth (in \%) for a charge 4 VC, and an aperture with 15\% central obscuration and 0.5\% thick spiders. The contrast is measured between the IWA and the OWA of the coronagraph, inside an annulus.}
\label{ChromaticContrast}
\centering
\end{figure}

Fig. \ref{ChromaticContrast} shows how the contrast changes with the bandwidth for an AVC designed for a $CO_{A}=15\%$, $t_{S}=0.5\%$ aperture. The indicated contrast is the mean contrast inside an annulus centered on the star, and with a 2.3\ld inner radius, and 26\ld outer radius. At $\lambda=\lambda_{0}$, the physical size of the mask corresponds to the 32\ld for which the mask is designed. 



The contrast goes down with the bandwidth. A $10^{-7}$ contrast is obtained with a 34\% bandwidth, a $10^{-8}$ contrast with an 18\% bandwidth, and a $10^{-9}$ contrast with a 2\% bandwidth.

While a 1.5$\times 10^{-8}$ contrast in a 20\% bandwidth might be enough for a ground-based instrument, additional constraints may have to be set on the Lyot plane intensity to increase the bandwidth and work with lower contrast. The electric field can be computed for two wavelengths instead of a single one, and to do that two mask sizes must be considered in the optimization problem. To make sure that high-contrast is obtained at the two extremities of a 20\% bandwidth, one could for instance compute the electric field for 32 and 25.6 $\lambda/D$ focal plane masks. Additional constraints may have to be set in-between these two extremum values.

\section{Manufacturing}\label{Manufacturing}

The apodizers displayed in this paper are not absolutely binary, but it has been shown in previous papers \citep{Carlotti2011, Carlotti2013} that solutions to these optimization problems tend to be binary, i.e, that the mask transmission can be rounded without affecting the contrast as long as enough points, say 512 in one quadrant of the pupil plane for a $10^{-9}$ contrast, are used to discretize the apodizer.

A binary apodizer can be manufactured as a quasi-achromatic component. The most common way to manufacture these devices uses photolithography. Several microdots apodizers \citep{Martinez2009, Sivaramakrishnan2010} have been manufactured with this technique for the coronagraphs of the gemini planet imager (GPI, \cite{Macintosh2007}) project, and the  spectro-polarimetric high-contrast exoplanet research (SPHERE, \cite{Beuzit2008}) project. 

Laboratory experiments conducted for the development of the optics of the SPICA coronagraphic instrument \citep{Enya2008} showed that transmissive binary apodizers could provide a 8$\times10^{-8}$ contrast in the visible. The substrate that is used can introduce chromatic wavefront errors, as well as internal reflections. The smallest features in the binary apodizers developed for SPICA are similar in size to those of the apodizers presented in this paper.

\begin{figure}[]
\centering
\includegraphics[width=0.85\textwidth]{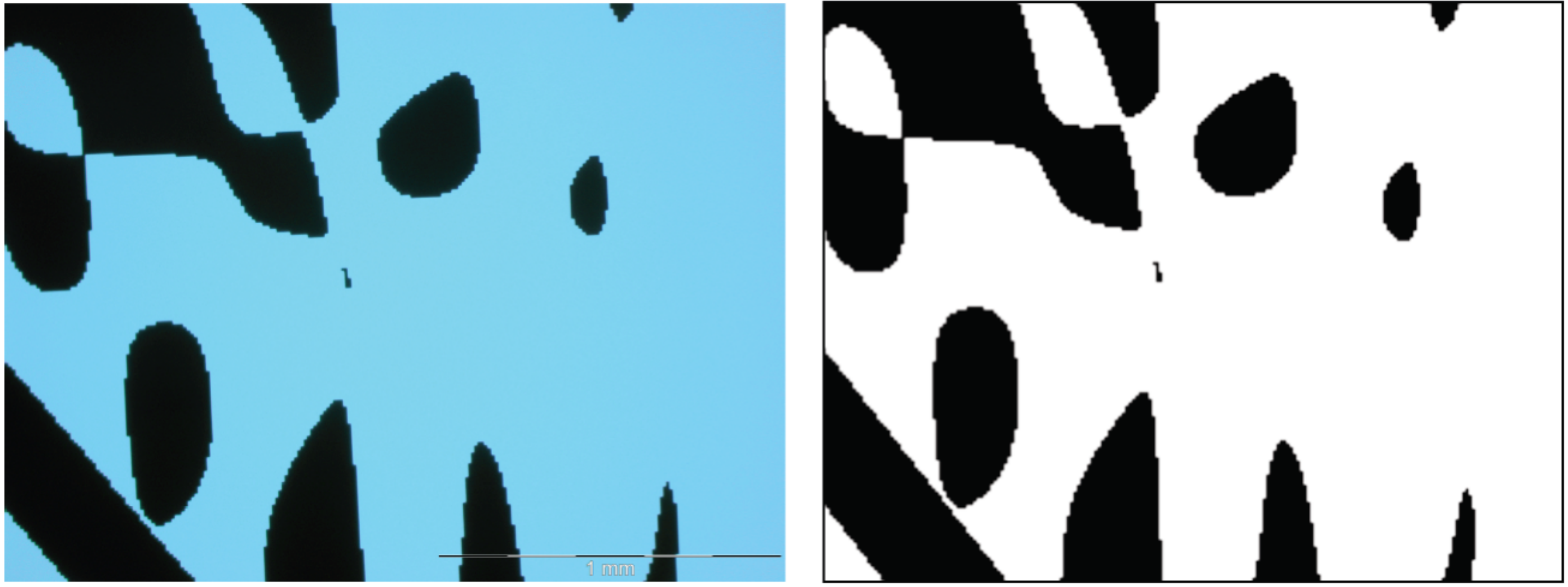}
\caption{Comparison of the theoretical and actual transmissions of a portion of a 2D shaped pupil manufactured for the Subaru telescope. Left: microscope photography of the mask. Right: corresponding desired profile. Each pixel has a 10$\mu m$ size.}
\label{CompManufacture}
\centering
\end{figure}

Several 2D shaped pupils have been manufactured in the last few years. The size of their smallest feature is virtually identical to that of the binary apodizers developed for SPICA. In particular two masks have been designed for the Subaru telescope, and they are now part of the SCExAO instrument (see \cite{Martinache2009,Carlotti2012spieSubaru}). Both have been manufactured using photolithography, with a resolution of about 10$\mu m$ (with about 1800 points across the apodizer). The high reliability of the etching process is illustrated in Fig. \ref{CompManufacture}. The smallest features are reproduced with fidelity, including 1-pixel large details.

Contrasts lower than $10^{-8}$ will require other types of binary apodizers, however. They must either be substrateless (and thus free-standing) or reflective.

The possibility of using MEMS or MOEMS devices such as micro-shutter arrays or micro-mirror arrays is particularly compelling at it would allow a dynamic control of the pupil plane amplitude, in addition to replacing several apodizers with a single device. This is especially interesting when considering the absence each night of several randomly selected mirrors in the pupil of the E-ELT (and potentially the TMT and other future telescopes with a large number of segments). Although it is not clear whether this technology is currently mature enough to be used for high-contrast imaging, such devices have been extensively tested for spectrophotometry instruments. Located in an image plane they allow the dynamical selection of multiple targets.

Deep reaction-ion etching - the same technique used to manufacture JWST-NIRSpec's micro-shutter arrays - can also be used to manufacture transmissive, freestanding masks. The ripple masks used in the high-contrast imaging laboratory at Princeton University, and in the high-contrast imaging testbed (HCIT) at the jet propulsion laboratory (JPL) were manufactured using this technique \citep{Belikov2007}. In this case small openings of a few tens of microns in size would be etched in a silicon wafer, with an orthonormal grid overlaying the mask to make it freestanding (and taken into account in the design process). Such masks would be achromatic, and would not introduce any wavefront aberration. Because of the nonzero width of the overlaying grid, the transmission of the mask would be slightly reduced, however.


Finally, reflective apodizers are currently in development at JPL \citep{Balasubramanian2013} in the context of the shaped pupil coronagraph developed for WFIRST-AFTA. A black silicon coating is deposited on a flat, reflective substrate to mask the nonreflective area of the apodizer.

\section{Discussion and conclusion}\label{Conclusion}

This paper builds upon and devises new methods to solve the limitations of two early papers:
\begin{itemize}
\item \cite{Mawet2013} presented 1D ring apodizers analytically optimized to create high-contrast with a VC and a 1D circular, centrally obscured aperture, without spiders.
\item \cite{Carlotti2013} presented 2D apodizers numerically optimized to create high-contrast with a 4QPM coronagraph and a 2D centrally obscured, arbitrarily shaped aperture, with spiders.
\end{itemize}

We have showed that apodizers can be numerically optimized in 2D to help restoring the high-contrast imaging capabilities of vortex coronagraphs when used with obscured apertures with arbitrary shapes and spiders. While the ring-apodizers have gray transmissions, these apodizers resemble the shaped pupils in that they have binary transmissions, and that they can be manufactured using the same processes.

Closed form expressions for even topological charges have been derived and used to directly compute the electric field in the Lyot plane as a function of the electric field in the first pupil plane. These pupil-to-pupil transforms have the advantage of not explicitly sampling the intermediate image plane and the vortex phase mask located there. One of the constraints of using these transforms is that the angular extent of the mask is limited. Given the computer currently used, it is possible to numerically optimize apodizers that are discretized over several hundred points on each axis of the pupil plane.

Because it is 2D, our formalism can be applied to the case of a segmented, noncircular aperture such as the pupil of one of the Keck telescopes. In practice the apodizer transmission is almost zero everywhere outside the inscribed circle defined by this noncircular aperture. Moreover, to maximize the transmission of the apodizer, the Lyot stop for which the apodizer is optimized must also be limited to that inscribed circle. This limitation most probably comes from the properties of the vortex mask, which is initially supposed to be used with a circular aperture. Apart from the apertures of the Keck telescopes, the GTC, or the JWST, only the apertures of the TMT and the E-ELT feature a noncircular outer edge, but their large number of segments make them almost circular, and this limitation does not affect them much.

Hence, we have considered circular obscured apertures with spiders. Apodizers have been computed for five different central obscuration ratios: 10, 15, 20, 25, and 30\%. For each of these obscurations, we have considered three different spider thicknesses (0, 0.5 and 1\%). The spiders form an orthogonal pattern. Apodizers have been optimized for charge $l=2$ and charge $l=4$ vortex phase masks. We were able to compute apodizers for 64 \ld OWA masks, but we have chosen to set the OWA to 32\ld since the substantially shorter time taken by the optimizations, and the smaller required amount of RAM, make it possible to solve several optimization problems at the same time, and thus explore a large number of different cases. This was necessary to exhaustively characterize the performance of the coronagraph in terms of throughput, IWA, and contrast.

In the optimization problems that we solve, the constraints are set on the intensity of the electric field computed using the pupil-to-pupil transform. We have chosen to set the attenuation of the on-axis light that goes through the Lyot stop to $10^{6}$. In practice this creates a $10^{-8}-10^{-10}$ contrast in the image plane depending on the transmission of the apodizer. The Lyot stops for which the coronagraphs have been designed have also been optimized, although their optimization has not been performed together with the optimization of the apodizers. Like all the apertures that we have considered, all Lyot stops are circular.

In the case of the spider-free apertures, the results of the numerical optimizations are similar to the ring-apodizers computed using the closed form expressions derived in \cite{Mawet2013}. Both types of apodizers are characterized by multiple rings with different mean transmissions. A similar aspect was found in our masks: the radii of their rings are close, and so are their transmissions.

For topological charges $l=2$, the throughputs are slightly smaller than those of the ring-apodized vortex coronagraphs. Like the charge $l=2$ vortex phase mask, the 4QPM is a $\theta^{2}$ coronagraph, and it is interesting to make  an indicative comparison between the apodizer computed for the 4QPM in \cite{Carlotti2013} for the aperture of the VLT telescope. The throughput of this optimal apodized 4QPM is 65\%, while the throughput of the optimal solution for an  AVC computed for the closest look alike aperture in the present paper (with a 15\% central obscuration, and 0.5\% thick spiders) is 46\%, which is substantially lower. This difference appears solely associated with the type of focal phase mask.

For topological charges of $l=4$, the throughputs of the optimal solutions are close to or larger than those of the RAVC, especially for large central obscurations. For instance, for a 20\% obscured aperture, the throughput of the numerically optimized apodized $l=4$ VC is 42\%, while it is about 32\% with the RAVC. For a 30\% obscuration, these throughput becomes 30\% and 13\%, respectively. We explain this difference by (a) the finite OWA of our vortex masks, and (b) the nonzero on-axis intensity that remains in the Lyot plane. The influence on the performance of the coronagraph of these two degrees of freedom has not been characterized yet, and doing so will be one of our next objectives.

Unlike the 1D apodizers showed in \cite{Mawet2013}, the 2D apodizers that we have presented take the aperture spiders into account. We chose to consider two nonzero thicknesses: 0.5 and 1\% of the diameter, the former being the thickness of the secondary supports of the 8m unit telescopes at the VLT. The throughput changes significantly with the spider thickness. While the throughputs associated with the 0.5\% thick spider apertures are - for both topological charges - about 3-10\% and 5-6\% smaller than those associated with spider-free apertures, the throughputs associated with the 1\% thick spider apertures are 9-33\% and 16-28\% smaller, respectively.

In addition to the horizontal and vertical alignments that are required by apodized coronagraphs, taking the spiders into account requires a clocking alignment as well. The spiders for which the apodizer is designed can be oversized with respect to the aperture spiders to accommodate clocking errors.

For small off-axis angles $\theta$, the transmission of off-axis sources follows a $\theta^{l}$ function, which confirms the fact that, like the VC, the AVC is an $l$-th order coronograph. The off-axis transmission of the AVC is not as monotonic as it is with the VC, however. The rate at which the transmission increases slows down, stalls or even decreases at around $\theta=1-1.2\ld$, and it only resumes its maximum value at around $\theta=1.5-2\ld$. The strength of this effect goes with the central obscuration.

For $l=2$, this does not affect the IWA much: it remains close to 0.9\ld, except for the 1\% thick spiders for which the IWA becomes 1.1-1.5\ld. The same is not true for $l=4$. In this case the IWA of the AVC is larger than it is with the VC, and it is mostly affected by the obscuration ratio. It is close to 1.75\ld for a 10\% obscuration, and increases to 2.5\ld for a 20\% obscuration, and to 3.2\ld for a 30\% obscuration. It should be noted, however, that it is difficult to describe the resolution properties of the coronagraph by solely referring to the usual definition of the IWA, i.e., the distance at which 50\% of the light of the off-axis light is transmitted. In particular the angle associated with a 40\% transmission - about 1.4\ld - is very similar to that of the VC for obscuration values up to 20\%.

The active compensation of aperture discontinuities (ACAD) presented in \cite{Pueyo2013} is a pupil mapping concept that uses two deformable mirrors to reduce the effective thickness of the spiders without loosing photons. ACAD could be used to implement the ring-apodizers presented in \cite{Mawet2013}. The apodizers presented in this paper are a compelling counterpart to ACAD, which could be affected by the malfunction of some of the DM actuators. In this view, they offer a more conservative solution at only a moderate cost in throughput. Because the spider thickness has an important influence on the throughput, apodizers designed for the VC could also take advantage of ACAD.

Although shaped pupils are binary apodizers - which are inherently achromatic - these apodizers are designed for a fixed mask radius measured in units of \ld, and the fixed physical radius of the mask causes the PSF at different wavelengths to be attenuated in different ways. The chromatic effects may be small enough in some cases. For instance a 1.5$\times 10^{-8}$ maximum contrast is obtained in a 20\% bandwidth for an aperture with a 15\% central obscuration and 0.5\% spiders (resembling that of the VLT's UT). To make sure that lower contrasts are obtained in a large bandwidth, apodizers will have to be optimized for multiple mask sizes.

\bibliographystyle{aa} 
\bibliography{BIB} 

\begin{thebibliography}{39}
\expandafter\ifx\csname natexlab\endcsname\relax\def\natexlab#1{#1}\fi

\bibitem[{{Balasubramanian} {et~al.}(2013){Balasubramanian}, {Wilson}, {White},
  {Muller}, {Dickie}, {Yee}, {Ruiz}, {Shaklan}, {Cady}, {Kern}, {Belikov},
  {Guyon}, \& {Kasdin}}]{Balasubramanian2013}
{Balasubramanian}, K., {Wilson}, D., {White}, V., {et~al.} 2013, in Society of
  Photo-Optical Instrumentation Engineers (SPIE) Conference Series, Vol. 8864,
  Society of Photo-Optical Instrumentation Engineers (SPIE) Conference Series

\bibitem[{{Belikov} {et~al.}(2007){Belikov}, {Give'on}, {Kern}, {Cady}, {Carr},
  {Shaklan}, {Balasubramanian}, {White}, {Echternach}, {Dickie}, {Trauger},
  {Kuhnert}, \& {Kasdin}}]{Belikov2007}
{Belikov}, R., {Give'on}, A., {Kern}, B., {et~al.} 2007, in Society of
  Photo-Optical Instrumentation Engineers (SPIE) Conference Series, Vol. 6693,
  Society of Photo-Optical Instrumentation Engineers (SPIE) Conference Series

\bibitem[{{Beuzit} {et~al.}(2008){Beuzit}, {Feldt}, {Dohlen}, {Mouillet},
  {Puget}, {Wildi}, {Abe}, {Antichi}, {Baruffolo}, {Baudoz}, {Boccaletti},
  {Carbillet}, {Charton}, {Claudi}, {Downing}, {Fabron}, {Feautrier},
  {Fedrigo}, {Fusco}, {Gach}, {Gratton}, {Henning}, {Hubin}, {Joos}, {Kasper},
  {Langlois}, {Lenzen}, {Moutou}, {Pavlov}, {Petit}, {Pragt}, {Rabou}, {Rigal},
  {Roelfsema}, {Rousset}, {Saisse}, {Schmid}, {Stadler}, {Thalmann}, {Turatto},
  {Udry}, {Vakili}, \& {Waters}}]{Beuzit2008}
{Beuzit}, J.-L., {Feldt}, M., {Dohlen}, K., {et~al.} 2008, in Society of
  Photo-Optical Instrumentation Engineers (SPIE) Conference Series, Vol. 7014,
  Society of Photo-Optical Instrumentation Engineers (SPIE) Conference Series

\bibitem[{{Carlotti}(2013)}]{Carlotti2013}
{Carlotti}, A. 2013, Astronomy \& Astrophysics, 551, A10

\bibitem[{{Carlotti} {et~al.}(2012){Carlotti}, {Kasdin}, {Martinache},
  {Vanderbei}, {Young}, {Che}, {Groff}, \& {Guyon}}]{Carlotti2012spieSubaru}
{Carlotti}, A., {Kasdin}, N.~J., {Martinache}, F., {et~al.} 2012, in Society of
  Photo-Optical Instrumentation Engineers (SPIE) Conference Series, Vol. 8446,
  Society of Photo-Optical Instrumentation Engineers (SPIE) Conference Series

\bibitem[{{Carlotti} {et~al.}(2013){Carlotti}, {Kasdin}, {Vanderbei}, \&
  {Riggs}}]{Carlotti2013SPIEc}
{Carlotti}, A., {Kasdin}, N.~J., {Vanderbei}, R.~J., \& {Riggs}, A. J.~E. 2013,
  in Society of Photo-Optical Instrumentation Engineers (SPIE) Conference
  Series, Society of Photo-Optical Instrumentation Engineers (SPIE) Conference
  Series

\bibitem[{{Carlotti} {et~al.}(2009){Carlotti}, {Ricort}, \&
  {Aime}}]{Carlotti2009}
{Carlotti}, A., {Ricort}, G., \& {Aime}, C. 2009, Astronomy and Astrophysics,
  504, 663

\bibitem[{Carlotti {et~al.}(2011)Carlotti, Vanderbei, \& Kasdin}]{Carlotti2011}
Carlotti, A., Vanderbei, R., \& Kasdin, N.~J. 2011, Opt. Express, 19, 26796

\bibitem[{{Enya} {et~al.}(2008){Enya}, {Abe}, {Tanaka}, {Nakagawa}, {Haze},
  {Sato}, \& {Wakayama}}]{Enya2008}
{Enya}, K., {Abe}, L., {Tanaka}, S., {et~al.} 2008, Astronomy \& Astrophysics,
  480, 899

\bibitem[{{Galicher} {et~al.}(2011){Galicher}, {Baudoz}, \&
  {Baudrand}}]{Galicher2011}
{Galicher}, R., {Baudoz}, P., \& {Baudrand}, J. 2011, Astronomy \&
  Astrophysics, 530, A43

\bibitem[{{Guyon}(2003)}]{Guyon2003}
{Guyon}, O. 2003, Astronomy \& Astrophysics, 404, 379

\bibitem[{{Guyon} {et~al.}(2010){Guyon}, {Martinache}, {Belikov}, \&
  {Soummer}}]{Guyon2010}
{Guyon}, O., {Martinache}, F., {Belikov}, R., \& {Soummer}, R. 2010, The
  Astrophysical Journal Supplement, 190, 220

\bibitem[{{Guyon} {et~al.}(2006){Guyon}, {Pluzhnik}, {Kuchner}, {Collins}, \&
  {Ridgway}}]{Guyon2006}
{Guyon}, O., {Pluzhnik}, E.~A., {Kuchner}, M.~J., {Collins}, B., \& {Ridgway},
  S.~T. 2006, The Astrophysical Journal Supplement Series, 167, 81

\bibitem[{{Kasdin} {et~al.}(2011){Kasdin}, {Carlotti}, {Pueyo}, {Groff}, \&
  {Vanderbei}}]{Kasdin2011}
{Kasdin}, N.~J., {Carlotti}, A., {Pueyo}, L., {Groff}, T., \& {Vanderbei}, R.
  2011, in Society of Photo-Optical Instrumentation Engineers (SPIE) Conference
  Series, Vol. 8151, Society of Photo-Optical Instrumentation Engineers (SPIE)
  Conference Series

\bibitem[{{Kasdin} {et~al.}(2007){Kasdin}, {Vanderbei}, \&
  {Belikov}}]{Kasdin2007}
{Kasdin}, N.~J., {Vanderbei}, R.~J., \& {Belikov}, R. 2007, Comptes Rendus
  Physique, 8, 312

\bibitem[{{Macintosh} {et~al.}(2007){Macintosh}, {Graham}, {Palmer}, {Doyon},
  {Gavel}, {Larkin}, {Oppenheimer}, {Saddlemyer}, {Wallace}, {Bauman},
  {Erikson}, {Poyneer}, {Sivaramakrishnan}, {Soummer}, \&
  {Veran}}]{Macintosh2007}
{Macintosh}, B., {Graham}, J., {Palmer}, D., {et~al.} 2007, Comptes Rendus
  Physique, 8, 365

\bibitem[{{Malbet} {et~al.}(1995){Malbet}, {Yu}, \& {Shao}}]{Malbet1995}
{Malbet}, F., {Yu}, J.~W., \& {Shao}, M. 1995, The Publications of the
  Astronomical Society of the Pacific, 107, 386

\bibitem[{{Martinache} \& {Guyon}(2009)}]{Martinache2009}
{Martinache}, F. \& {Guyon}, O. 2009, in Society of Photo-Optical
  Instrumentation Engineers (SPIE) Conference Series, Vol. 7440, Society of
  Photo-Optical Instrumentation Engineers (SPIE) Conference Series

\bibitem[{{Martinez} {et~al.}(2009){Martinez}, {Dorrer}, {Kasper},
  {Boccaletti}, \& {Dohlen}}]{Martinez2009}
{Martinez}, P., {Dorrer}, C., {Kasper}, M., {Boccaletti}, A., \& {Dohlen}, K.
  2009, Astronomy and Astrophysics, 500, 1281

\bibitem[{{Mawet} {et~al.}(2013{\natexlab{a}}){Mawet}, {Absil}, {Delacroix},
  {Girard}, {Milli}, {O'Neal}, {Baudoz}, {Boccaletti}, {Bourget},
  {Christiaens}, {Forsberg}, {Gonte}, {Habraken}, {Hanot}, {Karlsson},
  {Kasper}, {Lizon}, {Muzic}, {Olivier}, {Pe{\~n}a}, {Slusarenko},
  {Tacconi-Garman}, \& {Surdej}}]{Mawet2013VLT}
{Mawet}, D., {Absil}, O., {Delacroix}, C., {et~al.} 2013{\natexlab{a}},
  Astronomy and Astrophysics, 552, L13

\bibitem[{{Mawet} {et~al.}(2011{\natexlab{a}}){Mawet}, {Mennesson}, {Serabyn},
  {Stapelfeldt}, \& {Absil}}]{Mawet2011APJL}
{Mawet}, D., {Mennesson}, B., {Serabyn}, E., {Stapelfeldt}, K., \& {Absil}, O.
  2011{\natexlab{a}}, The Astrophysical Journal Letters, 738, L12

\bibitem[{{Mawet} {et~al.}(2013{\natexlab{b}}){Mawet}, {Pueyo}, {Carlotti},
  {Mennesson}, {Serabyn}, \& {Wallace}}]{Mawet2013}
{Mawet}, D., {Pueyo}, L., {Carlotti}, A., {et~al.} 2013{\natexlab{b}}, The
  Astrophysical Journal Supplement, 209, 7

\bibitem[{{Mawet} {et~al.}(2012){Mawet}, {Pueyo}, {Lawson}, {Mugnier}, {Traub},
  {Boccaletti}, {Trauger}, {Gladysz}, {Serabyn}, {Milli}, {Belikov}, {Kasper},
  {Baudoz}, {Macintosh}, {Marois}, {Oppenheimer}, {Barrett}, {Beuzit},
  {Devaney}, {Girard}, {Guyon}, {Krist}, {Mennesson}, {Mouillet}, {Murakami},
  {Poyneer}, {Savransky}, {V{\'e}rinaud}, \& {Wallace}}]{Mawet2012}
{Mawet}, D., {Pueyo}, L., {Lawson}, P., {et~al.} 2012, in Society of
  Photo-Optical Instrumentation Engineers (SPIE) Conference Series, Vol. 8442,
  Society of Photo-Optical Instrumentation Engineers (SPIE) Conference Series

\bibitem[{{Mawet} {et~al.}(2005){Mawet}, {Riaud}, {Absil}, {Baudrand}, \&
  {Surdej}}]{Mawet2005}
{Mawet}, D., {Riaud}, P., {Absil}, O., {Baudrand}, J., \& {Surdej}, J. 2005, in
  Society of Photo-Optical Instrumentation Engineers (SPIE) Conference Series,
  Vol. 5905, Society of Photo-Optical Instrumentation Engineers (SPIE)
  Conference Series, ed. D.~R. {Coulter}, 502--511

\bibitem[{{Mawet} {et~al.}(2011{\natexlab{b}}){Mawet}, {Serabyn}, {Wallace}, \&
  {Pueyo}}]{Mawet2011}
{Mawet}, D., {Serabyn}, E., {Wallace}, J.~K., \& {Pueyo}, L.
  2011{\natexlab{b}}, Optics Letters, 36, 1506

\bibitem[{{Murakami} {et~al.}(2008){Murakami}, {Uemura}, {Baba}, {Nishikawa},
  {Tamura}, {Hashimoto}, \& {Abe}}]{Murakami2008}
{Murakami}, N., {Uemura}, R., {Baba}, N., {et~al.} 2008, The Publications of
  the Astronomical Society of the Pacific, 120, 1112

\bibitem[{{Newman} {et~al.}(2013){Newman}, {Belikov}, \& {Guyon}}]{Newman2013}
{Newman}, K., {Belikov}, R., \& {Guyon}, O. 2013, in American Astronomical
  Society Meeting Abstracts, Vol. 221, American Astronomical Society Meeting
  Abstracts, 149.31

\bibitem[{{Pueyo} \& {Kasdin}(2007)}]{Pueyo2007}
{Pueyo}, L. \& {Kasdin}, N.~J. 2007, The Astrophysical Journal, 666, 609

\bibitem[{Pueyo {et~al.}(2009)Pueyo, Kay, Kasdin, Groff, McElwain, Give'on, \&
  Belikov}]{Pueyo2009}
Pueyo, L., Kay, J., Kasdin, N.~J., {et~al.} 2009, Appl. Opt., 48, 6296

\bibitem[{{Pueyo} \& {Norman}(2013)}]{Pueyo2013}
{Pueyo}, L. \& {Norman}, C. 2013, The Astrophysical Journal, 769, 102

\bibitem[{{Riggs} {et~al.}(2013){Riggs}, {Groff}, {Carlotti}, {Kasdin}, {Cady},
  {Kern}, \& {Kuhnert}}]{Riggs2013}
{Riggs}, A.~J.~E., {Groff}, T.~D., {Carlotti}, A., {et~al.} 2013, in Society of
  Photo-Optical Instrumentation Engineers (SPIE) Conference Series, Vol. 8864,
  Society of Photo-Optical Instrumentation Engineers (SPIE) Conference Series

\bibitem[{{Rouan} {et~al.}(2000){Rouan}, {Riaud}, {Boccaletti}, {Cl{\'e}net},
  \& {Labeyrie}}]{Rouan2000}
{Rouan}, D., {Riaud}, P., {Boccaletti}, A., {Cl{\'e}net}, Y., \& {Labeyrie}, A.
  2000, The Publications of the Astronomical Society of the Pacific, 112, 1479

\bibitem[{{Serabyn} {et~al.}(2010){Serabyn}, {Mawet}, \&
  {Burruss}}]{Serabyn2010Nat}
{Serabyn}, E., {Mawet}, D., \& {Burruss}, R. 2010, Nature, 464, 1018

\bibitem[{{Sivaramakrishnan} {et~al.}(2010){Sivaramakrishnan}, {Soummer},
  {Oppenheimer}, {Carr}, {Mey}, {Brenner}, {Mandeville}, {Zimmerman},
  {Macintosh}, {Graham}, {Saddlemyer}, {Bauman}, {Carlotti}, {Pueyo},
  {Tuthill}, {Dorrer}, {Roberts}, \& {Greenbaum}}]{Sivaramakrishnan2010}
{Sivaramakrishnan}, A., {Soummer}, R., {Oppenheimer}, B.~R., {et~al.} 2010, in
  Society of Photo-Optical Instrumentation Engineers (SPIE) Conference Series,
  Vol. 7735, Society of Photo-Optical Instrumentation Engineers (SPIE)
  Conference Series

\bibitem[{{Soummer} {et~al.}(2009){Soummer}, {Pueyo}, {Ferrari}, {Aime},
  {Sivaramakrishnan}, \& {Yaitskova}}]{Soummer2009Arbitrary}
{Soummer}, R., {Pueyo}, L., {Ferrari}, A., {et~al.} 2009, The Astrophysical
  Journal, 695, 695

\bibitem[{{Soummer} {et~al.}(2011){Soummer}, {Sivaramakrishnan}, {Pueyo},
  {Macintosh}, \& {Oppenheimer}}]{Soummer2011}
{Soummer}, R., {Sivaramakrishnan}, A., {Pueyo}, L., {Macintosh}, B., \&
  {Oppenheimer}, B.~R. 2011, The Astrophysical Journal, 729, 144

\bibitem[{{Spergel} \& {Kasdin}(2001)}]{Spergel2001}
{Spergel}, D. \& {Kasdin}, J. 2001, in Bulletin of the American Astronomical
  Society, Vol.~33, American Astronomical Society Meeting Abstracts, 1431

\bibitem[{{Vanderbei}(2012)}]{Vanderbei2012}
{Vanderbei}, R.~J. 2012, Mathematical Programming Computation, 590

\bibitem[{{Vanderbei} {et~al.}(2003){Vanderbei}, {Spergel}, \&
  {Kasdin}}]{Vanderbei2003}
{Vanderbei}, R.~J., {Spergel}, D.~N., \& {Kasdin}, N.~J. 2003, The
  Astrophysical Journal, 590, 593

\end{thebibliography}

\end{document}